\documentclass[fleqn,usenatbib]{mnras}
\usepackage[toc,page]{appendix}

\usepackage{newtxtext,newtxmath,nicefrac}
\usepackage{comment}

\usepackage[T1]{fontenc}

\DeclareRobustCommand{\VAN}[3]{#2}
\let\VANthebibliography\thebibliography
\def\thebibliography{\DeclareRobustCommand{\VAN}[3]{##3}\VANthebibliography}

\usepackage{graphicx}	
\usepackage{amsmath}	
\usepackage{orcidlink}

\title[Modified Conveyor Belt Model]{A Modified Conveyor Belt Model: Implications for Surface Density Thresholds for Massive Star Formation}

\author[Larose et al.]{
Nicholas Larose$^{1}$\orcidlink{0000-0001-2345-6789},
C. R. Kerton$^{1}$\orcidlink{0000-0003-1539-3321}, 
Kathryn Devine$^{2}$\orcidlink{0000-0002-3723-6362},
Grace Wolf-Chase$^{3}$ \orcidlink{0000-0002-9896-331X}
\\
$^{1}$Iowa State University, Department of Physics and Astronomy, 2323 Osborn Dr. Ames, IA 50011, USA\\
$^{2}$The College of Idaho, 2112 Cleveland Blvd., Caldwell, ID 8II5, USA\\
$^{3}$Planetary Science Institute, 1700 East Fort Lowell, Suite 106, Tucson, AZ 85719 USA
}

\date{Accepted XXX. Received YYY; in original form ZZZ}

\pubyear{2026}

\begin{document}

\graphicspath{{./}{figures/}}

\label{firstpage}
\pagerange{\pageref{firstpage}--\pageref{lastpage}}
\maketitle

\begin{abstract}
Recent models and simulations of cluster formation within molecular clumps consider multi-scale, hierarchical accretion, which leads to clump mass growth over time. This mode of mass accumulation could have implications regarding the evolution of observable properties such as mass and radius, bringing into question the interpretation of commonly cited thresholds for high-mass star formation. In this paper, we use the conveyor belt model of cluster formation to create synthetic cores/clumps and derive physical and observational properties. We show that while this model successfully predicts many observed trends, modifications are required to match properties of high-mass prestellar clumps. When the model clumps are observationally classified as intermediate- or high-mass star-forming, the threshold delineating these two groups agrees with those found in the literature; however, results show that high-mass clumps at early evolutionary stages can be misclassified using standard surface density thresholds. Our logistic regression analysis reveals the quantity of material to ever enter a star-forming region is the most important factor in differentiating intermediate- and high-mass star-forming regions. This implies observations characterising the environment surrounding star-forming regions are crucial, especially at early evolutionary stages. 

\end{abstract}


\section{Introduction} \label{sec:intro}
High-mass stars (those $\gtrsim \text{8-10} \;$M$_\odot$) are thought to dominate the energy budget of galaxies, trigger star formation, and provide the mechanisms to synthesise heavy elements through fusion and remnant mergers \citep{Motte2018}. The study of how these stars form, including the initial conditions for their formation, is therefore an important area of research. The birthplaces of high-mass stars are thought to be molecular clumps, which are parsec-scale condensations of dust and gas found within giant molecular clouds (\citealt{KrumholzMcKeeHawthorn2019,Beuther2025}). Large-scale surveys in the far-infrared and sub-millimetre such as Hi-GAL, the Herschel Infrared Galactic Plane Survey \citep{Molinari2010}, ATLASGAL, the APEX Telescope Large Area Survey of the Galaxy at 870 $\mu$m  \citep{Schuller2009}, and BGPS, the Bolocam Galactic Plane Survey \citep{Aguirre2011}, have resulted in the publication of large catalogues (e.g. \citealt{Svoboda2016,Elia2021,Urquhart2022}) containing data on clump properties at various stages of evolution. These catalogues provide invaluable data for testing models of high-mass star formation.

Observations and modelling suggest that star-forming clumps are not fully isolated from their surrounding environment and can accrete on multiple scales simultaneously. This is seen in models such as the global-hierarchical collapse model, \citep{Semadeni2019}, the conveyor belt model (\citealt{KrumholzMcKee2019,KrumholzMcKee2020}), and the hub-filament system model \citep{Kumar2020}. Evidence for this scenario appears in observations of BGPS clumps in various stages of evolution \citep{Svoboda2016} and in hub-filament systems \citep{Liu2023}, which both see trends in gas mass and surface density increasing as a function of their evolutionary stage. Simulations of clusters forming from molecular clouds (e.g. \citealt{Lee2016,Ibanez2017,Gonzalez2020,Camacho2020}) also show these trends.

Surface density thresholds are commonly used to identify regions of massive star formation (e.g. \citealt{Krumholz2008}). However the evolutionary trends identified above suggest that early stages of massive star formation can be misclassified, calling into question the use of a fixed surface density threshold. We note that \citet{Burkert2013} also challenged the necessity of fixed thresholds by demonstrating that observations originally implying their need (\citealt{Heiderman2010, Lada2010}) could be explained with multi-scale, continuous gas flows.

In this paper, we use analytic models of cluster formation that include mass accretion on multiple scales to investigate the properties of star-forming clumps as they evolve. In doing so, the following questions will be addressed: Can you distinguish between intermediate-mass and high-mass star-forming clumps early in their evolution? Does the surface-density threshold vary significantly as a function of evolutionary stage? Are there factors that control which clumps ultimately become high-mass star forming regions? 

An outline of the initial model used and the construction of synthetic clumps is presented in Section~\ref{sec:Modeling}. Section~\ref{sec:init_results} discusses the results and compares them with observations. A new model, developed to address issues present in the initial model, is explored in Section~\ref{sec:HMPreSClumps}. Section~\ref{sec:Analysis} compares results from the modified model with commonly used massive star formation thresholds, and observational properties of the synthetic clumps are analysed in an attempt to differentiate intermediate- and high-mass star forming regions. Our conclusions are presented in Section~\ref{sec:Conclusions}.

\section{Modelling}
\label{sec:Modeling}
Numerous unique models of high-mass star formation have been developed in an attempt to explain observational data. The Turbulent Core model \citep{McKeeTan2003} is an extension of the low-mass star formation paradigm for star or stellar system formation in cores. In this model, supersonically turbulent cores have high accretion rates forming massive stars on short time-scales ($\sim$ 10$^5$ yr). However, this model faces difficulties in explaining the lack of observed high-mass starless cores \citep{Motte2018}. It also assumes a fixed core mass, which is not consistent with the observations mentioned in Section~\ref{sec:intro} that suggest the core is not isolated from its surroundings. 

The global hierarchical collapse (GHC) scenario \citep{Semadeni2019} predicts initially compressed sheets collapsing into filaments, clumps, cores and finally stars, with low-mass stars forming over a much longer time-scale than more massive ones. This non-equilibrium model predicts collapse to occur at all scales, with each scale accreting from their parent structure. 

Similarly, by creating surface-density maps of line-of-sight filaments to nearby Hi-GAL sources, \citet{Kumar2020} proposed the `filaments to clusters' paradigm for star formation through observations of candidate hub-filament systems (HFSs). This idea stems around the coalescence of flow-driven filaments to form a junction called a hub, which is fed material from the inflowing filaments. Although the filaments may become dense enough at early times to produce low-mass stars, the accelerated accretion of material within the central hub is able to produce high-mass stars later on. This paradigm naturally produces age distributions seen in young star clusters, with much older low-mass stars in proximity to younger high-mass stars. 

The inertial-inflow model \citep{Padoan2020} challenged the core- and competitive-accretion models, finding that naturally occurring, large scale ($\gtrsim$ 1 pc), turbulent flows compress gas into filamentary structures that then go on to produce HFSs. This model predicts much longer time-scales ($\sim$1-6 Myr) for high-mass star formation than expected from other models and observations. 

Another alternative model, initially proposed by \citet{Longmore2014} and formalised by \citet{KrumholzMcKee2019} and \citet[][KM20 hereafter]{KrumholzMcKee2020}, is the conveyor belt model. This model considers mass inflow onto clumps with star formation occurring simultaneously, qualitatively sharing similarities with the HFS and GHC scenarios. One advantage of this model is that it is consistent with the lack of observations of gas clouds as massive and dense as the most dense star clusters (KM20).

Recently, KM20 compared their conveyor belt model to some of the aforementioned models, along with the static cloud scenario and the increasing star-formation efficiency model. To test the models, they performed a Markov chain Monte Carlo (MCMC) analysis on distributions of stellar ages in young clusters and ATLASGAL clump star-formation efficiency estimates, and compared the models' Galactic star formation budgets. They determined their conveyor belt model provided the best match to the observables.

Interestingly, \cite{Semadeni2024} determined that models explored in KM20 share many similarities with each other and their own GHC scenario. They stated that the GHC model explored in KM20 was oversimplified (assuming a monolithic collapse as opposed to a global, hierarchical one with simultaneous accretion at all scales) and concluded that the conveyor belt model better described the GHC scenario than the KM20 global-collapse model. Additionally, a conveyor belt mechanism would explain a growth in the observed physical properties (such as mass and surface density) of high-mass BGPS clumps, as discussed in \citet{Svoboda2016}. Thus we chose the conveyor belt model as the starting point for this work given that it shares many qualitative and quantitative similarities to competing models of high-mass star/cluster formation, and is able to produce distributions in agreement with observations.

\subsection{Initial Model}
In this section, we provide a brief summary of the conveyor belt model with rapid dispersal from KM20, as well as modifications tailored for the goals of this paper. The general time evolution of the gas and stellar mass for a star-forming clump has the form 
\begin{equation}\label{eq:GenFramework}
    \dot M_g = \dot M_{\text{acc}} - (1+\eta)\epsilon_{\text{ff}} \frac{M_g}{t_{\text{ff}}}, \; \; \; \; \dot M_* = \epsilon_{\text{ff}}\frac{M_g}{t_{\text{ff}}},
\end{equation}
where $\epsilon_{\text{ff}}$ is the star formation efficiency per free-fall time, $t_{\text{ff}}$ is the free-fall time, defined as $t_{\text{ff}} = \sqrt{3\pi / 32G\rho}$, $\eta$ is the mass loading factor (during accretion) which represents the ratio of the mass removal rate via feedback to the star formation rate (SFR), and $M_g$, $M_*$ and $\dot M_\text{acc}$ are the instantaneous gas mass, stellar mass, and accretion rate, respectively. The analytic form of the conveyor belt model is derived by assuming that the accretion rate accelerates until ceasing due to stellar feedback as
\begin{equation}\label{eq:AccRate}
    \dot M_\text{acc} = H(t_\text{acc}-t)(p+1)\frac{M_\text{g,0}}{t_\text{acc}} \Big( \frac{t}{t_\text{acc}} \Big)^p,
\end{equation}
where $M_\text{g,0}$ is defined as the total mass that will eventually reach the protocluster, $t_\text{acc}$ is the time over which accretion onto the star-forming region happens, $H(x)$ is the Heaviside step function, and \textit{p} is the exponent assuming the accretion rate accelerates as $t^p$. Solving equation~(\ref{eq:GenFramework}) using equation~(\ref{eq:AccRate}), and assuming rapid dispersal post-accretion yields
\begin{equation}
\label{eq:M*t}
    M_* = 
    \left\{
        \begin{aligned}
             & \frac{M_\text{g,0}}{(1+\eta)(p+2)\tau_\text{acc}^{p+1}}g(\tau,p+2) \quad & \tau \leq \tau_\text{acc} \\
             &  M_*(\tau_\text{acc}) + \frac{M_g(\tau_\text{acc})}{1+\eta_\text{d}}(1-e^{-\phi_\text{d}(\tau-\tau_\text{acc})}) \quad & \tau > \tau_\text{acc}
        \end{aligned}
    \right.
\end{equation}

\begin{equation}
\label{eq:Mgt}
    M_g = 
    \left\{
        \begin{aligned}
             & \frac{M_\text{g,0}}{\tau_\text{acc}^{p+1}}g(\tau,p+1) \quad & \tau \leq \tau_\text{acc} \\
             & M_g(\tau_\text{acc})e^{-\phi_\text{d}(\tau-\tau_\text{acc})} \quad & \tau > \tau_\text{acc} \ ,           
        \end{aligned}
    \right.
\end{equation}
where $\tau$ = $t/t_\text{sf}$, $t_\text{sf}$ is the star-formation timescale, $\tau_\text{acc}$ = $t_\text{acc}/t_\text{sf}$, $\eta_\text{d}$ is the mass loading factor (post-accretion), $\phi_\text{d} = (1+\eta_\text{d})/(1+\eta)$ determines how rapidly material is ejected from the star-forming region post-$\tau_\text{acc}$, and $g(\tau,p)$ is a monotonically increasing function of $\tau$, arising from the solution of equation~(\ref{eq:GenFramework}):
\begin{align}
    \label{eq:g}
    g(\tau, p) = p e^{-\tau} \int_0^\tau \tau'^{p-1} e^{\tau'}\, d\tau'.
\end{align}

\begin{figure}
    \centering
    \includegraphics[width=1\linewidth]{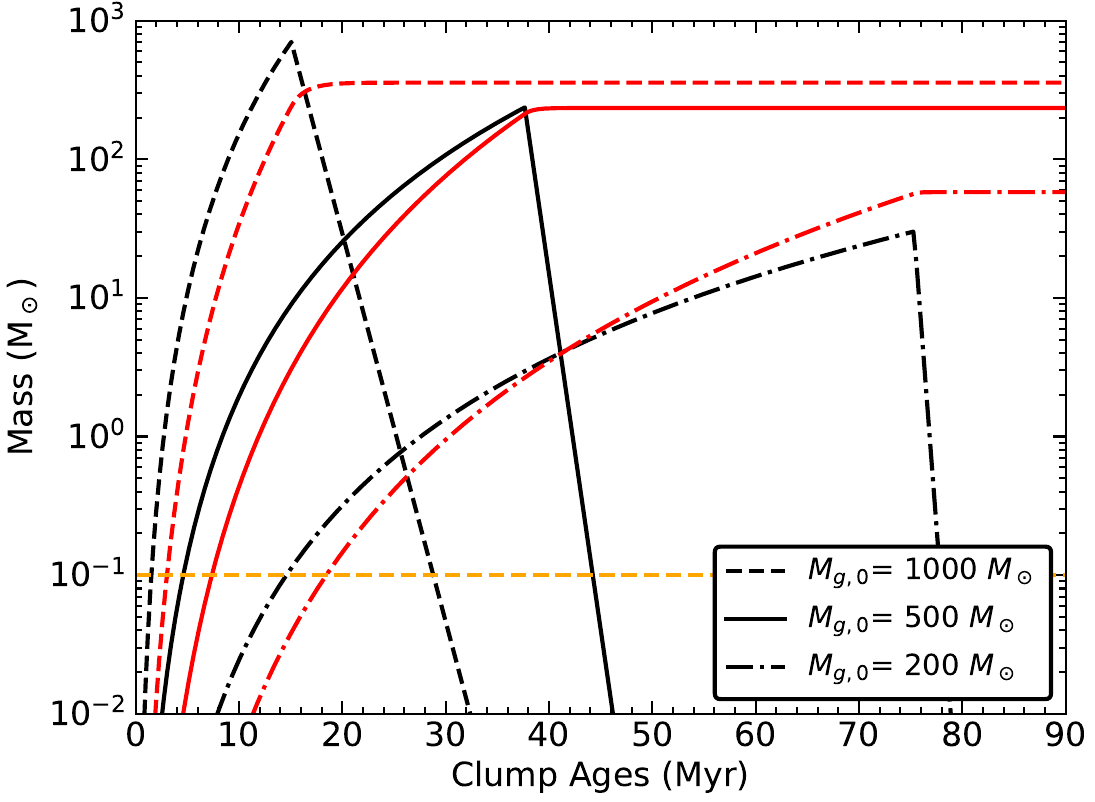}
    \caption{The evolution of the gas (black) and stellar (red) mass for the CBD model with representative values of $M_{g,0}$. These plots illustrate that in the CBD model the cloud mass is never fully assembled at a given time, and therefore does not initially need to be as massive as the final stellar mass. The orange dashed line indicates the minimum total stellar mass at which a model beings to populate a clump with YSOs (see text for details).}
    \label{fig:MgMsEvol}
\end{figure}

Through the remainder of this paper, the model described in this section will be referred to as the conveyor belt model with dispersal or `CBD'. The general evolution of the gas and stellar mass is seen in Fig.~\ref{fig:MgMsEvol} and is discussed in detail in Section~\ref{sec:init_results}.

\begin{table*}
    \centering
    \caption{Input parameters for the two models discussed in this paper, including their corresponding unit, range of input values and a short description. While more parameters are present in Section~\ref{sec:Modeling} and Section~\ref{sec:ModifiedModels} (e.g. $\eta_d$, $M_{{g,0}_\text{eff}}$), they can be derived from the parameters present here. Top: Table of input parameters for the CBD model. Bottom: Additional/modified input parameters for the Seeded CBD model, discussed in \autoref{sec:ModifiedModels}.}
    \label{tab:Parameter_Descriptions}
    \begin{tabular}{cccc}
        \hline
        Parameter & Unit & Range & Description \\
        \hline
        $M_\text{g,0}$  & M$_\odot$ & $10^2-10^5$ & Total amount of material that enters the star-forming region, including gas mass, stellar mass, and outflow material\\
        $t$ & yr & $10^4-10^9$ & Clump age \\
        $t_\text{sf}$ & yr & $1.2\times10^6-2\times10^7$ & Star formation timescale, $t_\text{sf} = t_\text{ff}(1+\eta)^{-1}\epsilon_\text{ff}^{-1}$ \\
        $t_\text{acc}$ & yr & $3.4\times10^6-1\times10^8$ & Accretion timescale \\ 
        $\eta$  & unitless & $10^{-2}-10^1$ & Mass loading factor = $\dot M_\text{fb}/\dot M_*$; the ratio of mass removal rate by feedback to the SFR\\
        \textit{p}  & unitless & 3 & Power law exponent for accretion rate, $\dot M_\text{acc}\sim \tau^p$\\
        $\epsilon_\text{ff}$ & unitless & 0.01 & Star formation efficiency per free-fall time \\
        $\phi_\text{d}$  & unitless & $10^{0.32}-10^{1.7}$ & Ratio of star-formation efficiencies before and after $\tau_\text{acc}$, equivalent to $(1+\eta_d)(1+\eta)^{-1}$\\
        \hline
        $M_\text{g,0}$  & M$_\odot$ & $10^2-10^{3.5}$ & Input parameter to control the total amount of material that ever enters the star-forming region, \\
        & & & which is larger by a factor of (1+$\beta$)$^{(p+1)}$\\
        $\beta$ & unitless & $10^{-1.92}-10^{0.64}$ & Number of accretion timescales prior to the onset of star formation, $\beta = \tau_0/\tau_\text{acc}$\\
        $t$ & yr & $10^0-10^9$ & Clump age \\
        $f_M$ & unitless & $10^{-5}-10^0$ & Fraction of mass that the model begins with in units of $M_\text{g,0}$ \\
        \hline
    \end{tabular}
\end{table*}

\subsection{Parameter Values}
KM20 fit their conveyor belt model to both the distribution of stellar ages in young clusters and the ATLASGAL $\epsilon_\text{ff}$ in protoclusters, producing ranges of parameters that agreed with both observational constraints simultaneously.

Both the $\eta$ and $\phi_\text{d}$ distributions appear uniform when inspecting their YSO posterior PDF for dimensionless parameters in their fig.~4, so both were uniformly sampled between their 16th and 84th percentile. While the distributions for $\phi_\text{d}$ derived from the distribution of stellar ages appear more constrained, to sample the largest number of model combinations the larger range from the $\epsilon_\text{ff}$ distribution was selected. KM20 initially assumes an arbitrary value of \textit{p}, and later concludes that $p=3$ holds the best agreement with observations. Visual inspection of posterior PDFs for $t_\text{sf}$ and $t_\text{acc}$ reveal non-uniform structure, so these distributions were sampled directly. These input parameters are summarized in Table~\ref{tab:Parameter_Descriptions}. 

\subsection{Advantages and Limitations}\label{sec:AdvLim}
The parametrized, analytic description of the CBD model allows for simple determination of the gas and stellar mass as a function of time. This in turn makes it possible to run thousands of models with different initial conditions allowing for large-scale statistical analysis, in contrast to using numerical integration or time consuming simulations. However, the benefits of the analytic model come at a cost, as many of the parameters treated as independent inputs require self-consistent solutions for improved physical accuracy. For instance, the accretion rate varies as a power law in the CBD model, but the analytic models of \citet{ZamoraAviles2025} found exponential growth at early stages when considering self-consistency between mass and radius. Additionally, models presented by \citet{ZamoraAviles2012} and \citet{Semadeni2024_delay} both considered the interdependence between the accretion rate and instantaneous gas mass.

The lack of self-consistency in the model parameter choices may therefore lead to some combinations which are not physically realistic. However, a primary goal of this work is to construct a statistically significant sample of cluster-forming clumps in order to explore the broadest range of parameter space possible. Future work would therefore require investigation into the physically plausible range of model parameter combinations.

\subsection{Synthetic Clump Construction}
While the model parameters provide gas and stellar masses over time, synthetic observations require populating the models with YSOs. To do this a similar procedure to that outlined in \citet[][M19 hereafter]{Molinari2019} was performed and is briefly summarised here. 

For each synthetic clump (i.e. set of model parameters) values for $M_g$ and $M_*$ are obtained from $10^4$ to $10^9$ years in 30 log-spaced intervals. The lower limit of $10^4$ yr is due to the fact that all clumps from the CBD model prior to this age fall outside of observability criteria (due to the fact that they are low-mass, low temperature greybodies) discussed later in Section~\ref{sec:init_results}, and would not be considered for further analysis. The upper limit of $10^9$ yr is selected as accretion timescales sampled from the KM20 distributions can exceed values of $\sim$100 Myr, and because star formation can briefly continue post-$\tau_\text{acc}$ depending on the value of $\phi_\text{d}$. For this work it is important to know the end state of star formation, that is whether or not the system eventually produces an intermediate- or high-mass star.

$M_{g,0}$ was chosen to range between $10^2$ and $\sim$$10^4$ M$_\odot$. While both lower and higher mass clumps could be created, smaller clumps are unlikely to form any significant stellar content and would not be useful in differentiating intermediate- versus high-mass star-forming regions. Larger values of $M_{g,0}$ were excluded as above roughly $10^{3.5}~M_\odot$ all models eventually produced high-mass stars. 

However, combinations of $M_\text{g,0}$ with $t_\text{acc}$ may not be physically meaningful. For example, a large value of $M_\text{g,0}$ combined with a small $t_\text{acc}$ may be associated with unphysical star-formation timescales. Exploring this in detail would require identification of physically plausible parameter combinations, and would be best approached with self-consistent models, such as those discussed in \autoref{sec:AdvLim}.

The clumps themselves are assumed to initially comprise low-temperature, optically thin gas and dust. The opacity is taken to be $\kappa_{300}$ = $0.027$ cm$^2$ g$^{-1}$ (per gram of gas+dust), assuming the $R_V$=5.5 dust model from \citet{Draine2003}. The temperature of a star-forming clump is assumed to increase with time due to intra-cluster heating from protostars. In the Hi-GAL Compact Source Catalogue-II \citep[][CSC-II hereafter]{Elia2021}, the temperature distribution of prestellar and protostellar compact sources, as well as candidate \ion{H}{II} regions are determined. 

The compact source dust temperatures increase with evolutionary stage from $11.4\,{\rm K}$ (prestellar) to $15.2\,{\rm K}$ (protostellar) and $24.5\,{\rm K}$ K (\ion{H}{II} region candidates) in the inner galaxy. Similarly in the outer galaxy, the dust temperature increases from $10.5\,{\rm K}$, to $15.3\,{\rm K}$ and $23.9\,{\rm K}$ for pre-, protostellar and \ion{H}{II} region candidates respectively. In our model, starting dust temperatures are randomly sampled from the Hi-GAL prestellar temperature distribution. The final dust temperature is taken from either the protostellar or \ion{H}{II} region candidate temperature distributions. If the final state of the model has a YSO with mass $>$ 8 M$_\odot$, the \ion{H}{II} region candidate distribution is sampled, leading to potentially higher final temperatures, otherwise the protostellar distribution is sampled. Given that the intra-clump heating is primarily caused by embedded YSOs, the rate that the temperature increases is chosen to follow the SFR. While the dust emission from the clump will contribute to the total flux, the dust will also cause extinction for the embedded YSOs, requiring prior information on the clump's size. As mentioned in KM20 the radii of clumps as they evolve could be static, increasing or even decreasing. 

One of the major assumptions for the CBD model was that $t_\text{ff}$ is constant throughout the evolution of a given model. As discussed in section 2.2.2 of KM20, simulations and models suggest that as long as the accretion rate is high enough, the molecular cloud growth time is comparable to its free-fall time. This results in a density and SFR per free-fall time that does not evolve significantly with time. Because the gas mass for the CBD model is known from equation~(\ref{eq:Mgt}), the free-fall time can be inverted to determine the free-fall radius ($R_\text{ff} = (8GMt_\text{ff}^2/\pi^2)^{1/3}$) as a proxy, allowing the clump to grow for $\tau\leq\tau_\text{acc}$. 

In order to populate the parent clump, the total stellar mass at each time-step is used to sample a Kroupa initial mass function (IMF) \citep{Kroupa2001} between 0.1 and 50 M$_\odot$. It is assumed that the total number of YSOs either remains constant or increases with time. These cases respectively represent mass accretion onto protostars or the creation of new ones. The \citet[][R06 hereafter]{Robitaille2006} model SEDs are then associated with each YSO. This is done by searching for the shortest Euclidean distance in (log($M_*$), log($t$)) space, where the age is determined through random sampling between the youngest age in the R06 models ($10^3$ yr) and the time since star formation began in the clump. If the total stellar mass is beneath the limit to form stars (in this case 0.1 M$_\odot$), the clump is not populated and is assumed to be prestellar. Random line-of-sight orientations are simulated by choosing inclinations of the closest R06 spectral energy distribution (SED) model between 0 and 90$^\circ$. 

The physical properties of the R06 YSOs allow them to be classified as zero-age main sequence (ZAMS) if they meet criteria specific to main sequence stars. To accomplish this, information from the $\text{[Fe/H]} = +0.00$ and $v/v_\text{crit}$=0.4 MIST isochrone models \citep{Choi2016} is used, where $v_\text{crit}$ is the critical surface linear velocity. These are models with non-zero rotation, which are more realistic than the alternative $v/v_\text{crit}$=0.0 models due to the fact that higher mass stars exhibit significant rotation rates. Candidate high-mass ZAMS stars are defined as those which meet the mass, age, and temperature criteria for MIST B3-O9 V stars, which are then flagged for later classification as described in Section~\ref{sec:Analysis}. Stars with properties signifying an earlier spectral classification than O9 V ($M\sim18\,{\rm M_\odot}$, $t\sim 10^{4.7}$ yr, and $T_\text{eff}\sim34~000\,{\rm K}$) continue to fall into the O9 category, but simply represent higher mass main-sequence stars. 

If the combination of mass, age, and temperature is between that of a A1 V ($M\sim2\,{\rm M_\odot}$, $t\sim 10^7$yr, and $T_\text{eff}\sim9200\,{\rm K}$) and a B3 V ($M\sim5.6\,{\rm M_\odot}$, $t\sim 10^{5.9}$yr, and $T_\text{eff}\sim18~000\,{\rm K}$) star, it is classified as an intermediate-mass star. Those which can be classified as ZAMS have only the stellar component of the R06 SED extracted (excluding, for instance, any remaining envelope). Because the CBD models evolve over significantly long time-scales, it is possible that some of the embedded high-mass stars exceed their MS lifetimes. The maximum age of the MIST isochrones are recorded and interpolated so that if a randomly sampled YSO age exceeds its MS lifetime, it is flagged so that it may be removed from further consideration. This removes models which would have likely undergone a supernova explosion or some other post-MS event which would rapidly disrupt the clump's natal environment. 

\begin{figure}
    \centering
    \includegraphics[width=1.0\linewidth]{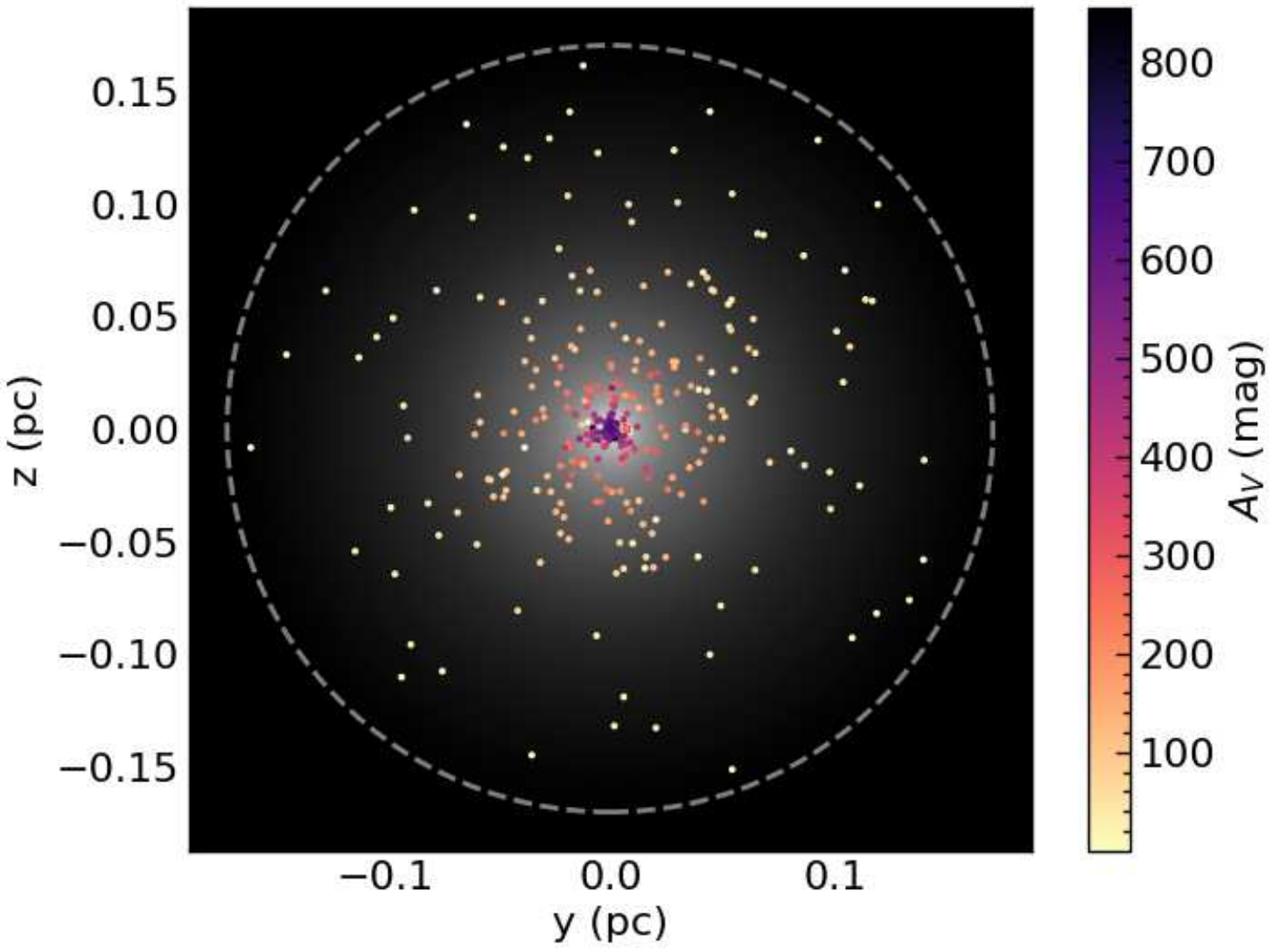}
    \caption{Example of an evolved clump populated with YSOs. The dots represent the YSO positions as a projection on the y-z plane, and extinction is calculated along the x-axis towards the `observer'. The background gradient represents the density profile of the clump, while the YSOs are colour-coded by their visual extinction (see text for details).}
    \label{fig:StellarDist}
\end{figure}

Following M19, we randomly chose YSO locations in a 3-dimensional space within the clump radius ($R_\text{ff}$), using a source radial density profile with a power-law exponent of $-2.5$. These YSOs are embedded within a clump with a dust radial density profile of $\rho = \rho_0 [r/r_0]^{-1.5}$, where $r_0=0.01$~pc and $\rho_0$ is assumed constant within $r_0$. Note the source power-law and dust radial density exponents were chosen to match those used in M19. These profiles place YSOs preferentially towards the centre of the clump in a higher-extinction environment. An example of a synthetic clump with the locations of the YSOs can be seen in Fig.~\ref{fig:StellarDist}. 

\begin{figure}
    \centering
    \includegraphics[width=1.0\linewidth]{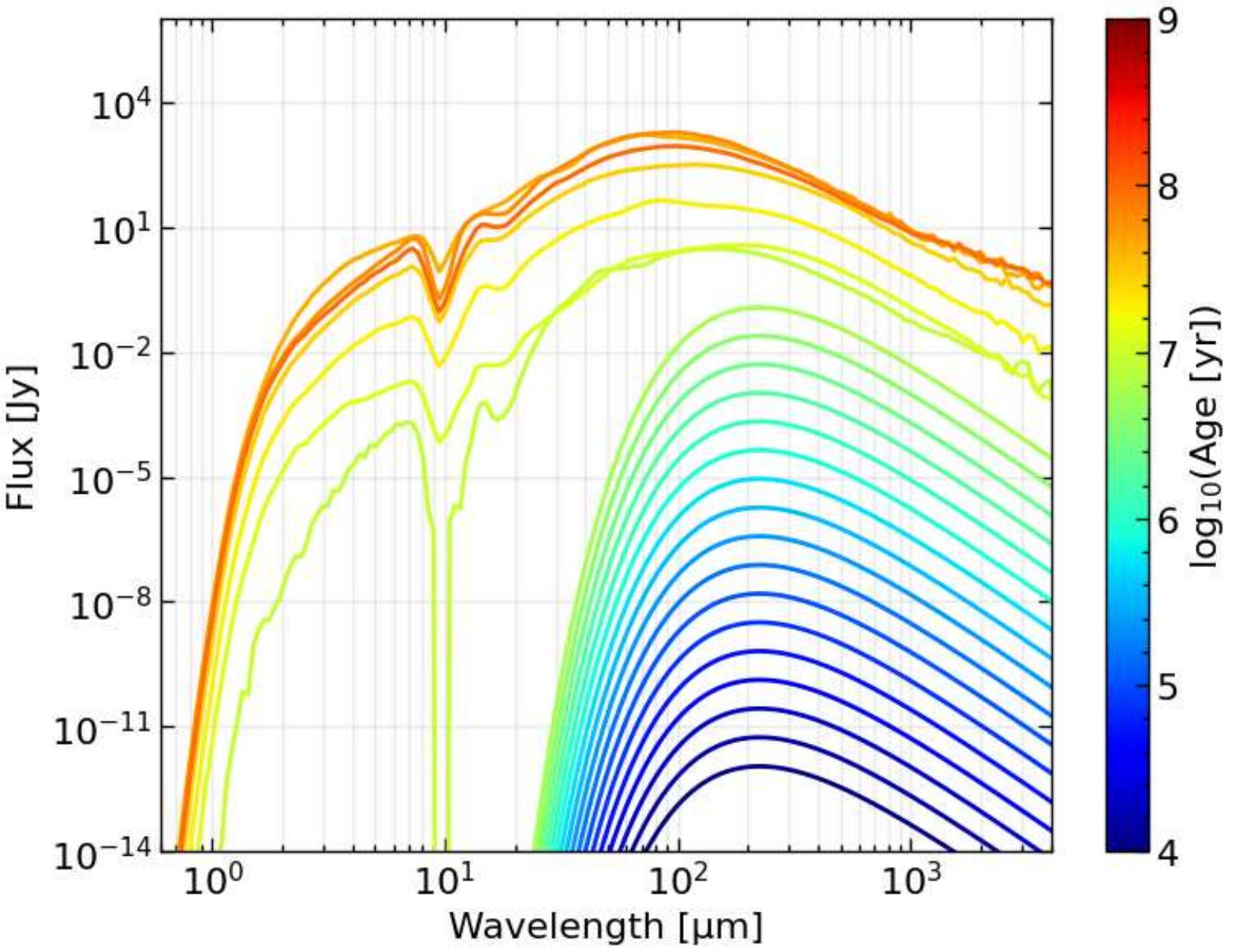}
    \caption{Example SED evolution of a clump from birth to the end of the model run. The SEDs are colour coded by age. This model spends $\sim$7 Myr in the prestellar phase before enough material is accumulated for YSOs to populate the clump, causing a sharp jump in luminosity, and emitting significant flux at shorter wavelengths.}
    \label{fig:SEDEvol}
\end{figure}

The total SED is constructed by summing the SEDs of every YSO (adjusted for dust extinction) and the dust emission from the parent clump. This total SED is then integrated to determine the luminosity of the clump at each time step. To account for the variation in IMF and age sampling of the sources, each set of input parameters is run twenty times, with median properties (such as luminosity and SED shape) being reported. An example of this can be seen in Fig.~\ref{fig:SEDEvol}.

To summarise, from a set of input parameters the CBD model yields the clump gas and stellar mass at 30 log-spaced timesteps between $10^4$-$10^9$ yr. The mass values at each timestep are used to populate the greybody modelled clump with YSOs and produce a total SED. This process is repeated 20 times to account for variations in the IMF and age sampling, where the final set of observational properties are given by the median values. A total of $\sim5\times10^3$ models were created, with 30 time-steps per model resulting in $\sim1.5\times10^5$ model clumps at various stages of evolution.

\section{Comparison of CBD to Observations}
\label{sec:init_results}
Provided the complete set of model parameters and SEDs, the synthetic clumps can be compared to observations. Many surveys have determined the physical properties of clumps in different stages of evolution (see Section~\ref{sec:intro}); however, this study will use the Hi-GAL CSC-II catalogue as the primary dataset for comparison as it has the largest collection of mid-IR selected compact sources/clumps in the literature. Their sample of approximately $1.5\times10^5$ objects spans a wide range of evolutionary stages with the luminosity to mass ratio (L/M), a common proxy for age, ranging from $\sim$0.001 to $>10^4$ L$_\odot$/M$_\odot$. 

\subsection{Data Selection and Constraints}
Before comparing the model clump properties with observations, those with flux densities not detectable by \textit{Herschel} and/or not meeting CSC-II catalogue inclusion criteria were removed from our sample. We applied the high-reliability criteria from \citet{Elia2017} that requires three or more adjacent (non-zero) flux measurements at $160$, $250$, $350$, and $500~\mu$m. To limit arbitrarily small fluxes, the instrument noise limits used for the Hi-GAL dataset were adopted as a lower limit for observability. The Hi-GAL observing mode results in point-source instrument noise values of 45.2, 12.1 10.0 and 14.4~mJy~beam$^{-1}$ for PACS 160 $\mu$m, and SPIRE 250, 350 and $500~\mu$m respectively \citep{herschel_parallel_mode_2011}. These constraints preferentially remove very early stages of the model clump evolution, which is modelled as a low-temperature, low-mass greybody. Throughout the remainder of the paper references to the CBD models will refer to this subset. 

\subsection{Gas Mass, Stellar Mass vs. Time}
Shown in Fig.~\ref{fig:MgMsEvol} are the gas and stellar masses as a function of time for representative models. The models use the median sampled $t_\text{sf}$ of 7.5$\times$10$^{6}$ yr and $\eta$=0.23, $\tau_\text{acc}$= 2, 5, 10 and $\eta_d$= 5, 10, 20 corresponding to models with $M_\text{g,0}$ of 1000, 500 and 200 M$_\odot$ respectively.

Due to the large value of \textit{p}, accretion begins slowly but ramps up as $\tau$ approaches $\tau_\text{acc}$. This can lead to the instantaneous stellar mass exceeding the gas mass pre-$\tau_\text{acc}$ (as seen in the $M_\text{g,0}\sim$200~M$_\odot$ model in Fig.~\ref{fig:MgMsEvol}). Accretion timescales range between 6.5 and 105 Myr with the majority (61 per cent) lying within  62$\pm$23 Myr. The range of randomly sampled $M_\text{g,0}$ values results in gas masses ranging between $\sim10^{-2}$ and $\sim10^4$ M$_\odot$. The majority (87~per cent) of maximum gas masses obtained for each model fall within log($M_\text{max}$) = $2.1 \pm 0.7 \; $M$_\odot$, the largest being $1.5\times10^4$ M$_\odot$. While reference to larger clumps can be found in the literature, over 99.5~per cent of clumps in the CSC-II catalogue lie beneath $10^4$ M$_\odot$. The large median value of $\phi_d$ ($\sim$ 10.5) causes the gas mass post-$\tau_\text{acc}$ to be expelled quickly with star formation ceasing shortly after. 

The horizontal line at 0.1 M$_\odot$ in Fig.~\ref{fig:MgMsEvol} indicates the minimum stellar mass required to populate the clump with YSOs. The range of clump gas masses where stars begin to form, and thus become protostellar, is between $0.1$ and $\sim30$ M$_\odot$, which severely limits the mass corresponding to the prestellar phase of star formation. This will be further discussed in the next section and Section~\ref{sec:Analysis}. 

\subsection{L vs. M}
\begin{figure*}
    \centering
    \includegraphics[width=1\linewidth]{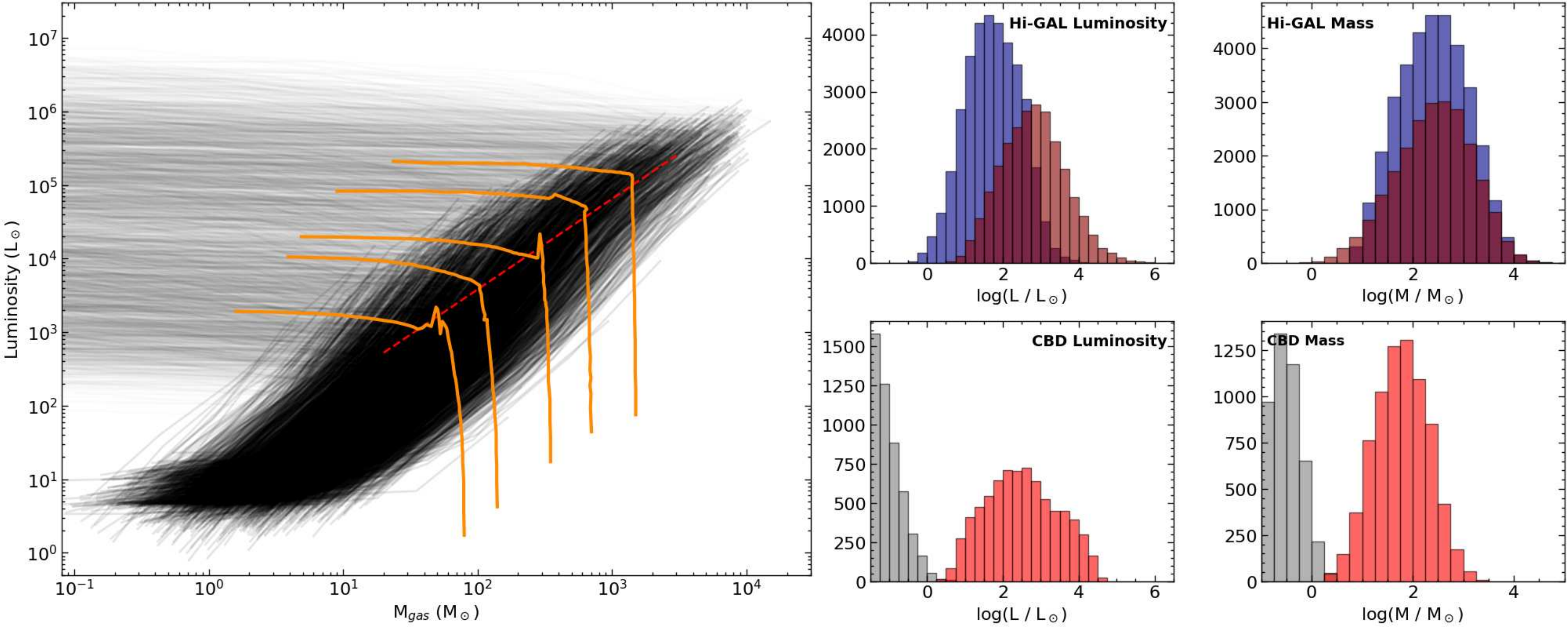}
    \caption{\textit{Left}: Evolutionary tracks for the accreting protostellar portion of the CBD model. The post-$\tau_\text{acc}$ phase is shown at a higher transparency in the background as to not take away from the structure. The prestellar phase is omitted from the evolutionary tracks as they begin at arbitrarily small values and increase up to the lowest protostellar points shown around $\sim1-10$ L$_\odot$. The evolutionary tracks from M08 are overlaid in orange. The red dashed line indicates the average location of a model at $\tau_\text{acc}$, or where the transition from an accretion dominated phase to an `envelope clean up' phase begins. This is consistent with the M08 transition point, where their evolutionary tracks shift from vertical to horizontal. \textit{Right}: Histograms comparing the prestellar (blue) and protostellar (red) distributions for Hi-GAL clumps and CBD model clumps (i.e., those with radii between 0.1 and 1.5 pc), with a 0.25 bin width. The lack of prestellar points in the CBD distributions is due to the fact that none of the models become large enough to be classified as clumps before star formation begins. Instead gray histograms show where all non-clump prestellar CBD models lie.
    \label{fig:LM_origmodel}}
\end{figure*}

The left-hand panel of Fig.~\ref{fig:LM_origmodel} shows the luminosity versus mass evolutionary tracks for every CBD model once star formation has begun. Overlaid are the original \citet{Molinari2008} evolutionary tracks for pre-assembled clumps forming single massive YSOs. During the star-formation phase, their tracks are roughly vertical, with increasing clump luminosities and roughly constant gas masses. At $\tau_\text{acc}$ accretion stops. Physically this arises from different forms of stellar feedback. At lower masses this includes jets and outflows, while at higher masses, photodissociation, photoionization, and stellar winds become more important \citep{Beuther2025}.

This is followed by a gas-dispersal phase where the luminosity of the clump primarily reflects the stellar content and the gas mass decreases as the clump is dispersed. M19 modified this picture to include cluster formation on a larger scale with varying fractions of starting material corresponding to compact cores, leading to gas mass decreasing at a constant clump-to-core accretion rate, but the general shape of the tracks remained unchanged. In contrast, the conveyor belt model exhibits much different behaviour. In this case, the models accumulate both gas and stellar material following $\dot M\sim t^3$, and therefore traverse the LM plane at an angle during the star-formation phase. 

The right-hand panel of Fig.~\ref{fig:LM_origmodel} shows histograms of prestellar (blue) and protostellar (red) clumps (i.e. those with radii between $0.1$ and $1.5$ pc) in comparison with clumps from the Hi-GAL CSC-II catalogue. The prestellar and protostellar clumps for the CBD model are defined as follows; the prestellar clumps are those without any YSOs, or equivalently a total stellar mass $< 0.1 $ M$_\odot$, while the protostellar clumps have at least one YSO or $\geq 0.1 $ M$_\odot$ of total stellar mass. An alternative way to observationally define model prestellar and protostellar clumps is discussed in \hyperref[appendix:alt_evol_class]{Appendix~\ref*{appendix:alt_evol_class}}.  

The lack of the highest mass CBD clumps in comparison to those from the CSC-II is simply due to the parameter range spanned in this paper. As referenced in Section~\ref{sec:Modeling}, the maximum $M_\text{g,0}$ value sampled for the CBD model is $3\times10^4$ M$_\odot$. While larger values would in turn create more massive protostellar clumps, they are not required for comparison of models which do and do not create high-mass stars. 

There is however an absence of prestellar CBD clumps over the same mass and luminosity range as the CSC-II clumps. This does not represent an issue with the CBD model, but is instead due to the fact that the model, by construction, does not include a prestellar stage. The reason for the lack of prestellar clumps is that none of the models are able to reach large enough radii before forming stars and thus becoming protostellar. If all prestellar sources were included regardless of size (gray histogram defined as `non-clumps' in Fig.~\ref{fig:LM_origmodel}), the significant disagreement with the observed prestellar CSC-II masses and luminosities would be seen.

Essentially, for the CBD model, no combinations of model parameters are able to produce high-mass prestellar clumps when using either of our definitions of prestellar. To illustrate this, the gas to stellar mass ratio can be used to show that the CBD model (with the range of parameters from KM20) inherently has difficulty in producing high-mass prestellar clumps. This ratio using equation~(\ref{eq:M*t}) and equation~(\ref{eq:Mgt}) can be written as
\begin{equation}
        \frac{M_g}{M_*} = (p+2)(1+\eta)\frac{g(\tau, p+1)}{g(\tau, p+2)},
\end{equation}
where again $p=3$. The parameter $\eta$ can be used to determine the stellar mass at a given time (provided the gas mass is known), where larger values provide more gas mass at a given stellar mass. Larger values of $\eta$ allow for larger prestellar clumps (larger $M_g/M_*$), so the upper limit of log($\eta$)$\sim$0.26 from KM20 is adopted. For instance, if one wanted to produce a 500 M$_\odot$ prestellar clump (assuming the total stellar content at this point is $\approx0.1$ M$_\odot$), using the approximation for g($\tau$,p)$\approx\tau^p/(1+\tau/p)$ from KM20, this would need to occur prior to $\tau\approx0.003$. Given the range of plausible $t_\text{sf}$ values (5.6-16 Myr), this would require instantaneous accretion rates to exceed 0.04-0.13~M$_\odot$~yr$^{-1}$, which are in excess of those seen in massive HFSs. Additionally, if one were to assume a minium accretion timescale of $\tau_\text{acc}=1$, this would suggest an $M_{g,0}$ of $\sim8\times10^{12}$ M$_\odot$. Such clumps are significantly more massive than those seen in catalogues such as Hi-GAL CSC-II and are not physically plausible. This again demonstrates that the CBD model was not intended for the creation of prestellar clumps. Based on this, we developed a modification to the CBD model allowing for the existence of prestellar clumps, which is discussed in Section~\ref{sec:HMPreSClumps}.

\begin{figure}
    \centering
    \includegraphics[width=1\linewidth]{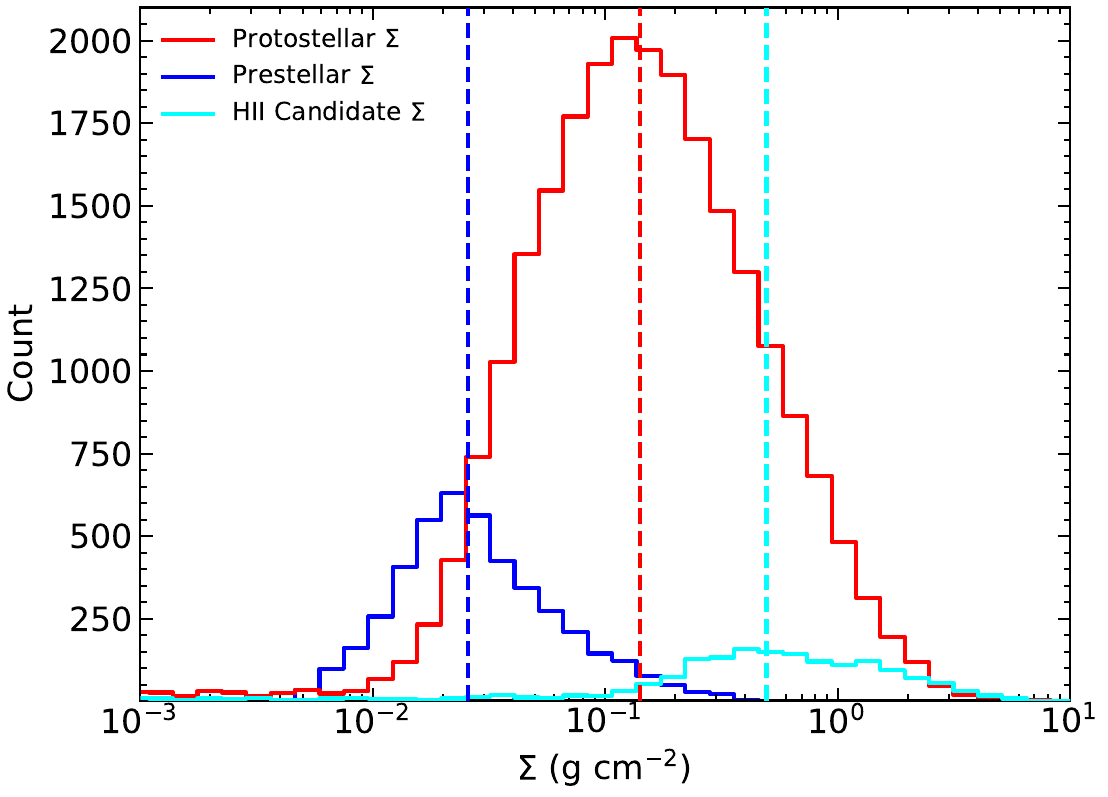}
    \caption{Surface density distribution for the CBD model. Points are defined as follows: \ion{H}{II} region candidates are those with ZAMS, high-mass stars (B3-O9 or earlier). Protostellar clumps are those with a total stellar mass above 0.1 M$_\odot$, and prestellar are those with less than 0.1 M$_\odot$.}
    \label{fig:SigmaDist}
\end{figure}

\subsection{Surface Density Distributions}
During the mass accretion phase we can define the surface density ($\Sigma$) of the CBD models as $M_g$/$\pi R_\text{ff}^2$. Fig.~\ref{fig:SigmaDist} shows the distribution of surface densities for prestellar and protostellar models. A new category, \ion{H}{II} region candidates, is introduced, which is defined as regions containing high-mass ZAMS stars (see Section~\ref{sec:Modeling}). We note that an observational threshold of $L/M\geq22.4$~L$_\odot$/M$_\odot$ has been used in the CSC-II catalogue to identify \ion{H}{II} region candidates. We show in Section~\ref{sec:Analysis} that the two definitions agree well. 

The median surface densities of the prestellar, protostellar and \ion{H}{II} region candidates are 0.026, 0.141, and 0.494 g cm$^{-2}$ respectively. This demonstrates an evolutionary trend in the increase of observed $\Sigma$ with time in the early stages of clump formation discussed in Section~\ref{sec:intro}. The assumptions of constant free-fall time and density allow the radius to be written as $R\sim(Mt_\text{ff}^2)^{1/3}$, resulting in the surface density scaling as $\Sigma \sim M^{1/3}$. The low-mass, prestellar models therefore reside only at low surface densities. After $\tau_\text{acc}$, the surface densities decrease as mass is lost due to stellar feedback and the models retain the same radius (see discussion below). It is interesting that the \ion{H}{II} region candidates have a wide range of surface densities. The spread is due to two factors: first, the \ion{H}{II} region candidates form near or after $\tau_\text{acc}$ so their surface density distribution peaks higher than both the prestellar and protostellar stages, and second, the long left-hand tail reflects more evolved \ion{H}{II} region candidates, which have undergone extensive mass loss.

The size of the clump becomes difficult to define post $\tau_\text{acc}$ due to various forms of feedback destroying the natal environment. The definition of the clump itself becomes ambiguous as well, given the models at these late time-steps are more akin to young clusters. Although it is likely that the `radius' would continue to increase from outflows and feedback, together with the cessation of accretion, the radius is held constant for simplicity.

\section{Seeded Conveyor Belt Model}
\label{sec:HMPreSClumps}
As mentioned in Section~\ref{sec:init_results}, while the CBD model was successfully able to reproduce the Galactic SFR, ATLASGAL $\epsilon_\text{ff}$ and young cluster age distributions simultaneously, it remains limited by the absence of a prestellar stage. Because stars begin to populate the clumps at unrealistically low masses only, this increases the total luminosities earlier than expected, resulting in evolutionary tracks missing a large fraction of the parameter space shared by clumps in the Hi-GAL CSC-II catalogue. In this section, an alternative model is presented which addresses this issue. 

\subsection{Model Description}
\label{sec:ModifiedModels}
In an attempt to reconcile the lack of high-mass prestellar clumps, while maintaining the results from KM20, a modified version of the CBD model is introduced in which the gaseous clump is allowed to form \textit{prior} to the onset of star formation. This raises the clumps' mass at $\tau=0$ so that higher mass prestellar clumps can exist. The model maintains the conveyor belt mechanism and assumes the accretion rate follows $\dot M_\text{acc}\sim t^3$. 
The gas and stellar mass as a function of time is obtained by re-solving equation~(\ref{eq:GenFramework}) and equation~(\ref{eq:AccRate}), while modifying the boundary condition such that the gas mass begins at some non-zero value, $M_\text{init}$. This value is randomly, uniformly sampled to be some fraction ($\log f_M\in$ [${-5}$,0]) of the total amount of material that enters the star-forming region ($M_\text{g,0}$). This range was chosen as values greater than one would imply a majority of the initially present clump material is present prior to the onset of star formation, which is more akin to the static cloud model (KM20) than the conveyor belt model. The lower limit allows for the randomly sampled values of $M_{g,0}$ ($\sim 10^2-10^3$ M$_\odot$) to result in $M_\text{init}$ ranging between $10^{-3}$ and $10^3$ M$_\odot$, which covers the range of prestellar clumps seen in both the Hi-GAL CSC-II and ATLASGAL catalogues well. 

It is important to note that some values of $f_M$ may be unrealistic, as the initial prestellar mass should be self-consistently determined by the accretion rate and initial clump radius, and not as a free parameter (see discussion in \autoref{sec:AdvLim}). As a result, some sampled values of $f_M$ could represent unrealistic combinations of $M_{g,0}$ and $M_\text{init}$, and would not be consistent with the other clump properties. While different ranges or distributions of $f_M$ may be more realistic than those chosen for this work (see results of MCMC fit in \autoref{sec:SCBD_Compared_to_obs}, with preference for larger values of $f_M$), we emphasize that $f_M$ is used as an exploratory prior, where again a more physically accurate prescription would require self-consistent solutions.

While this model shares many similarities to the CBD model, there are important observational differences. For instance, the gas mass contribution now comes from the initial `seeded' mass as well as the accreted material. Conceptually, this model can be thought of as a `delayed star-formation' CBD model, where the gas mass could begin at zero, then increase at zero star-formation efficiency for a time. Afterwards the region would become either dense or massive enough for multi-scale collapse to occur. The stellar mass is still proportional to the amount of gas mass present at a given time as seen in equation~(\ref{eq:GenFramework}), and because there can be a significantly large amount of gas mass at $\tau=0$, the final stellar mass (for a given set of input parameters) is generally larger than in the CBD model. The resulting gas and stellar masses under these assumptions are:
\begin{equation}
\label{eq:Mgtinittau0}
    M_g = 
    \left\{
        \begin{aligned}
             &M_\text{init} e^{-\tau}\\
             &\quad+\frac{M_\text{g,0}}{\tau_\text{acc}^{p+1}} 
             \Big[
                g(\tau+\tau_0, p+1) 
                - g(\tau_0, p+1) e^{-\tau}
            \Big] \quad &  \tau < \tau_\text{acc}\\
            &M_g(\tau_\text{acc})e^{-\phi_d(\tau-\tau_\text{acc})} \quad & \tau \geq \tau_\text{acc}    
        \end{aligned}
    \right.
\end{equation}

\begin{equation}
\label{eq:Msinittau0}
    M_* = 
    \left\{
        \begin{aligned}
             &\frac{1}{1+\eta} \Bigg[
                    M_\text{init}(1 - e^{-\tau})\\
                    &\quad + \frac{M_{g,0}}{\tau_\text{acc}^{p+1}} \Bigg(
                    \frac{g(\tau+\tau_0, p+2)}{p+2}
                    - \frac{g(\tau_0, p+2)}{p+2}\\
                    &\qquad - g(\tau, p+1)(1 - e^{-\tau})
                    \Bigg)
                    \Bigg] \quad & \tau < \tau_\text{acc} \\
            &M_*(\tau_\text{acc}) + \frac{1}{1+\eta} [M_g(\tau_\text{acc})(1-e^{-\phi_d(\tau-\tau_\text{acc}))})] \quad & \tau \geq \tau_\text{acc}
        \end{aligned}
    \right.
\end{equation}
where the function $g$ is the same as in equation~(\ref{eq:g}). A detailed derivation of these equations is presented in \hyperref[appendix:deriv_SCBD]{Appendix~\ref*{appendix:deriv_SCBD}}, and will be referenced hereafter as the `Seeded' CBD model, or SCBD model.

If one were to blindly implement the initial condition for a non-zero gas mass, there would be tension between the material being converted to stars and the small accretion rates. Near $\tau\approx0$, the accelerating accretion rate forces more gas mass to be lost to star formation than gained from the remainder of $M_{g,0}$, meaning the gas mass initially decreases up to hundreds of solar masses before accretion catches up, which then causes the gas mass to increase until $\tau_\text{acc}$. Since this behaviour is not seen in models of star formation (see Section~\ref{sec:intro}), the following condition to force the gas mass to monotonically increase was implemented. If the accretion begins due to the accumulation of material prior to the onset of star formation over some amount of time $\tau_0=\beta\tau_\text{acc}$, $\beta$ must be greater than $(\tau_\text{acc}f_M /(1+p))^{-p}$ to force $\dot M_\text{acc}$ to be positive at $\tau=0$. 

While $M_\text{g,0}$ represents the total amount of material to ever reach the protocluster in the CBD model, the addition of the seeded mass (and seed formation timescale $\tau_0$) in the SCBD model increases the accretion rate at $\tau\sim0$. Therefore the total amount of material to ever enter the protocluster (from $\tau=-\tau_0$ to $\tau=\tau_\text{acc}$) in the SCBD model is actually given by $M_\text{g,0}$(1+$\beta$)$^{(p+1)}$.

\begin{figure}
    \centering
    \includegraphics[width=0.8\linewidth]{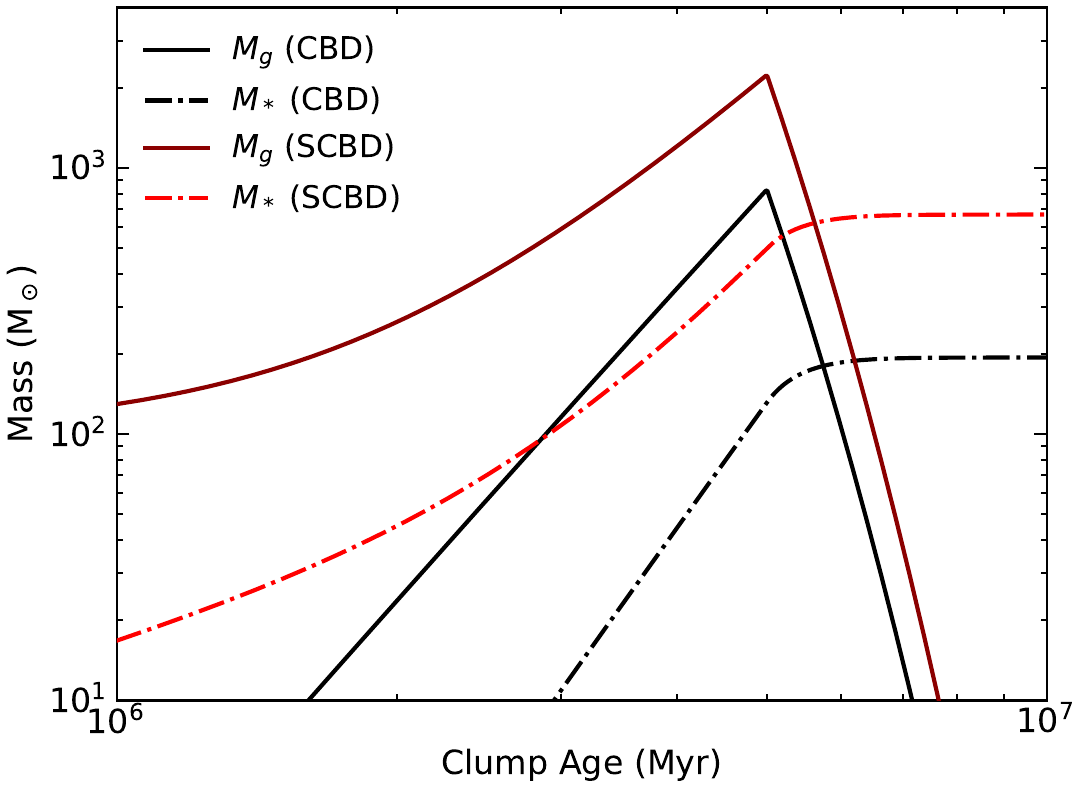}
    \caption{Gas (solid line) and stellar mass (dot-dashed line) evolution of the CBD and SCBD models, given identical input parameters (see text for details).}
    \label{fig:MgMsSeeded}
\end{figure}

Figure~\ref{fig:MgMsSeeded} illustrates the gas and stellar mass evolution assuming $M_{g,0}=10^3\,{\rm M_\odot}$ for both the CBD and SCBD models. The SCBD model has $M_\text{init}=0.1\,{M_\text{g,0}}$ with all other parameters identical ($t_\text{sf}$ = $5\times10^6$ yr, $\tau_\text{acc}=1$, $\eta = 0.3$, $\eta_d = 12.26$). We note that the $t_0$ value for the SCBD model shown is $\sim 2.3$ Myr. Larger $\tau_\text{acc}$ values naturally lead to increased $t_0$ values, for instance $\tau_\text{acc}=30$ results in $t_0\sim 20 \;$Myr. 

\subsection{SCBD Compared to CBD}
While the SCBD model includes prestellar stages ranging from low- to high-mass, to ensure the model is physically plausible it was tested using the publicly available MCMC software provided by KM20. To be viable, the SCBD model must reproduce the results discussed in Section~\ref{sec:Modeling} from KM20; it should be able to fit the stellar age distributions of Orion and NGC 6530, the $\epsilon_\text{ff}$ distribution from ATLASGAL, both individually and simultaneously, and finally predict a Galactic SFR beneath the maximum of $\sim$ 2 M$_\odot \; \text{yr}^{-1}$. 

\begin{figure}
    \centering
    \includegraphics[width=1\linewidth]{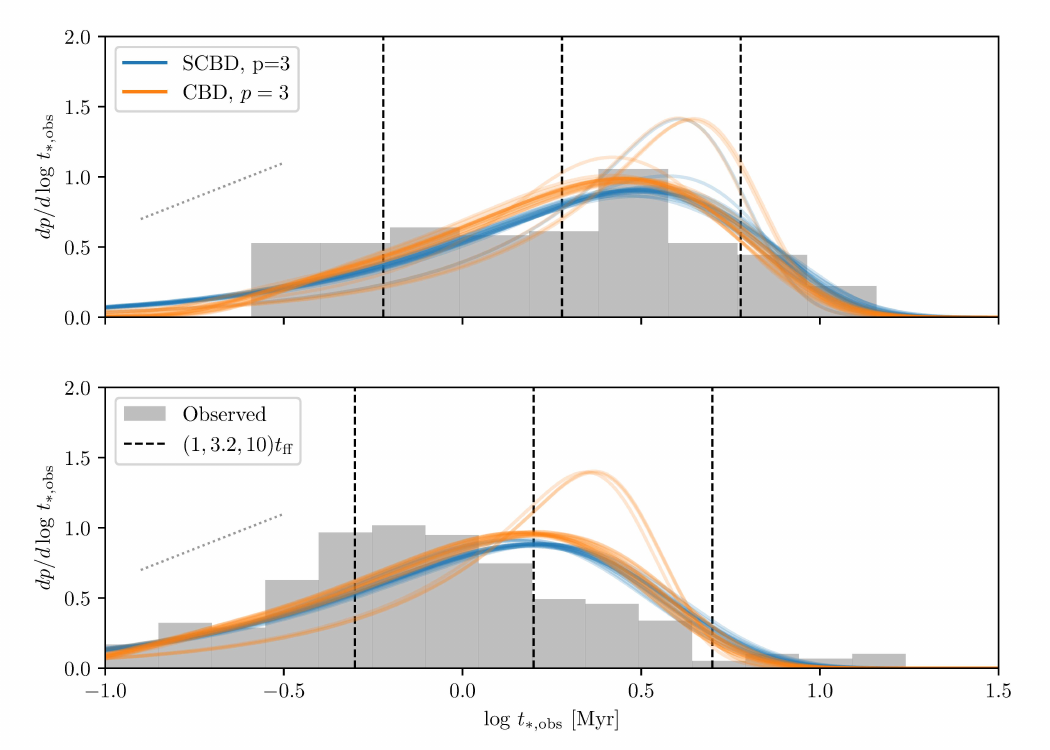}
    \caption{Age distributions of YSOs in Orion (top) and NGC 6530 (bottom) for the original CBD compared to the modified SCBD. The coloured lines represent 20 random samples from the final MCMC iteration for each model. Figure produced using the publicly available software from \citet{KrumholzMcKee2020}.}
    \label{fig:age_fits}
\end{figure}

Prior to the analysis of the stellar age distributions of  Orion and NGC 6530, varying values of \textit{p} were tested, as was done in KM20. While not shown here, $p=3$ also produced the best results in comparison to the young clusters and was therefore adopted for this model. Seen in Fig.~\ref{fig:age_fits} are the Orion and NGC 6530 stellar age distributions for both the CBD and SCBD models, both of which perform adequately in terms of matching the observed age distributions. The SCBD $\epsilon_\text{ff}$ distributions from ATLASGAL in Fig.~\ref{fig:eff_dist} also shows exceptional agreement with both the CBD and parent distributions. 

We then simultaneously fit the $\epsilon_\text{ff}$ and age distributions using the SCBD models (see Fig.~\ref{fig:combinedpdf}). This figure represents the posterior probability-density function (PDF) for the dimensionless parameters of the SCBD model ($\epsilon_\text{ff}$, $\eta$, $\tau_\text{acc}$, $\phi_\text{d}$, and additionally $f_\text{M}=M_\text{init}/M_\text{g,0}$), which are derived using the stellar age distribution from NGC 6530 (red) and YSO counts from ATLASGAL (blue). 

Similar to the distributions in the posterior PDF of the CBD model (fig. 4 in KM20), we see significant overlap for the parameters shared between the CBD and SCBD models. Additionally, the $\epsilon_\text{ff}$ and age distributions for the $f_M$ PDF seen in the lower right panel of Fig.~\ref{fig:combinedpdf} overlap. The degree of overlap for all of the parameters indicates that the SCBD model is able to simultaneously reproduce both $\epsilon_\text{ff}$ and age distributions. This is in contrast to the increasing efficiency (IE) model shown in fig. 6 of KM20, where half of the parameters share no common values.

Finally, the Galactic SFR was calculated to be $\sim 0.55 \;\text{M}_\odot \;\text{yr}^{-1}$, well within the Milky Way total of roughly $\sim2$ M$_\odot$ yr$^{-2}$. Given that this model mirrors the original CBD in many ways, it should come as no surprise that this model succeeds equally well when compared to the results from KM20. 
\begin{figure}
    \label{fig:eff_dist}
    \centering
    \includegraphics[width=1\linewidth]{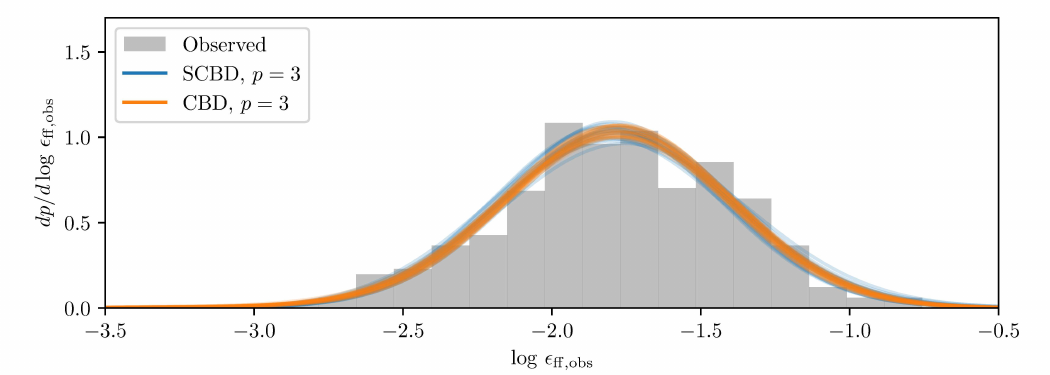}
    \caption{Distribution of $\epsilon_\text{ff}$ values of the original CBD compared to the modified SCBD. The coloured lines represent 20 random samples from the final MCMC iteration for each model. Figure produced using the publicly available software from \citet{KrumholzMcKee2020}.}
    \label{fig:eff_dist}
\end{figure}

\begin{figure}
    \centering
    \includegraphics[width=1\linewidth]{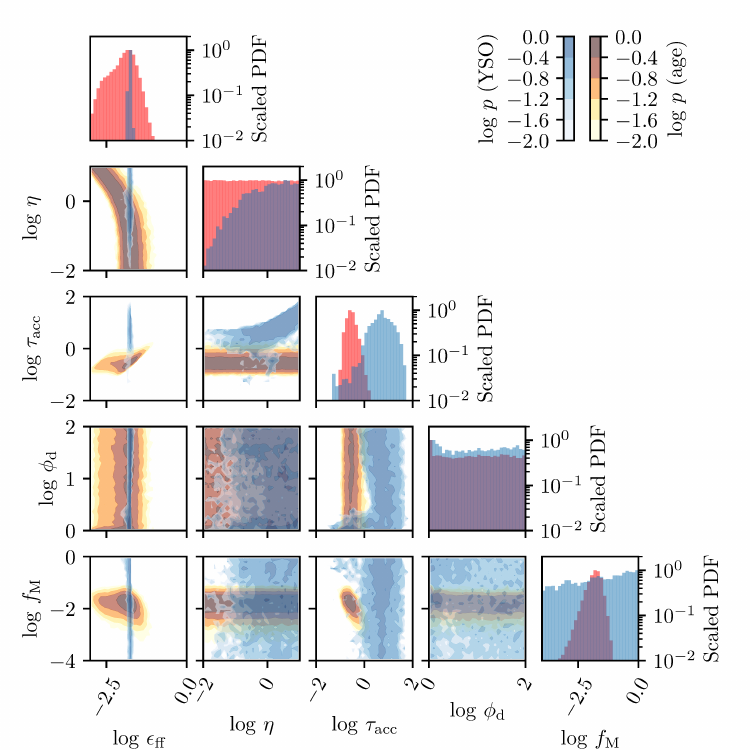}
    \caption{Combined PDF for simultaneous fits of ATLASGAL $\epsilon_\text{ff}$ and young cluster YSO age distributions for the SCBD model. The panels along the diagonal show the probability distributions for each parameter individually. The off-diagonal panels show how pairs of variables are related, with contour lines indicating the joint distributions. The blue colours represent posterior PDFs derived from YSO counts, while the red / orange colours represent the posterior PDFs derived from stellar ages in NGC 6530. Figure produced using the publicly available software from \citet{KrumholzMcKee2020}.}
    \label{fig:combinedpdf}
\end{figure}

\subsection{SCBD Compared to Observations}
\label{sec:SCBD_Compared_to_obs}
The range of physically acceptable parameters returned for the SCBD model is similar to those from the CBD. Additions to the input parameters are provided in the bottom panel of \autoref{tab:Parameter_Descriptions}. The same procedure as in Section~\ref{sec:Modeling}  can then be performed to compare the SCBD model outputs to observations. Note that in this case, the pre-assembled clumps at $\tau=0$ have sufficient mass to pass the observability criteria discussed in Section~\ref{sec:init_results}, therefore the time-steps used for the SCBD model range instead from $\tau=0$ to $\tau=\tau_\text{acc}$. As before, the SCBD model will refer to the reduced dataset which passes the same observability criteria for the CBD model.

\begin{figure*}
    \centering
    \includegraphics[width=1\linewidth]{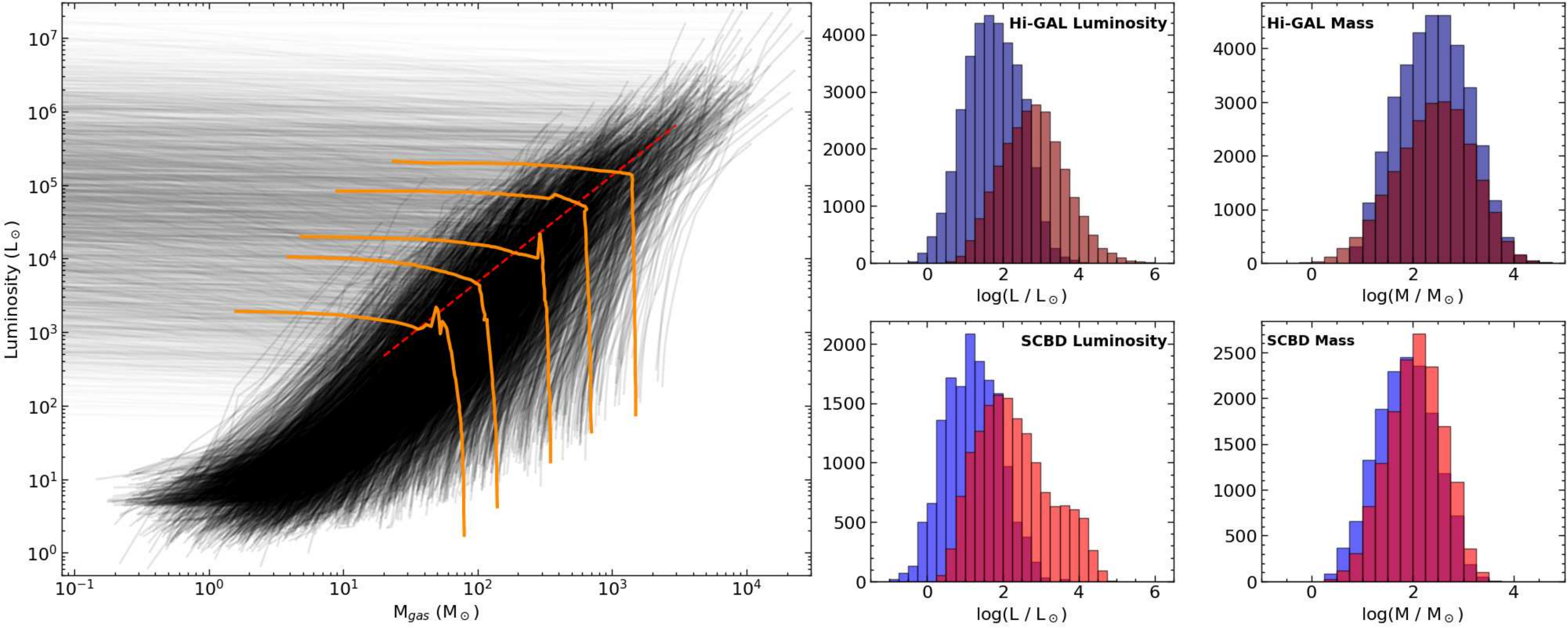}
    \caption{Same as in Fig.~\ref{fig:LM_origmodel}, except for the SCBD models.}
    \label{fig:MinitLvsM}
\end{figure*}

The left panel of Fig.~\ref{fig:MinitLvsM}  shows the resulting luminosity versus mass evolutionary tracks for the SCBD models once star-formation has begun. The tracks themselves are similarly skewed to the right relative to the \citet{Molinari2008} tracks; however, models with large values of $f_M$ tend to begin their evolution more vertically. In these cases, the SCBD model begins to share similarities with the static cloud model outlined in KM20, or the constant clump-to-core accretion rate model used in M19. This is due to the accretion rates being negligible near $\tau\sim0$, keeping the gas mass nearly constant while star formation causes a rise in luminosity. The red dashed line shows the best fit to the locations of $\tau_\text{acc}$, which represents the average location where the clumps stop accreting and enter the `envelope clean up phase'. 

One of the most obvious differences between the results of the SCBD and CBD models is seen in the right hand panel of Fig.~\ref{fig:MinitLvsM}. The large starting masses of the SCBD clumps prior to the onset of star formation shift the prestellar distributions further to the right, mirroring the distributions seen in the Hi-GAL CSC-II much better than in Fig.~\ref{fig:LM_origmodel}. 

We note that larger accretion rates are required to attain a monotonic gas mass increase once star formation begins. This leads to effective $M_{g,0}$ values larger by a factor of (1$+\beta$)$^{(p+1)}$, with $>$50 per cent of the SCBD models having an effective $M_{g,0}>10^{3.2}$ M$_\odot$. This results in a majority ($>$80 per cent) of models going on to produce high-mass stars, showing that although the range of input parameters was similar between the CBD and SCBD models, the latter can produce more massive star-forming regions with smaller values of $M_{g,0}$. 

As discussed in \citet{Elia2021}, \citet{Molinari2016} defined a threshold of 10 L$_\odot$/M$_\odot$ to identify clumps hosting main sequence stars. \citet{Elia2021} found that $\sim$22.5 per cent of their protostellar sources were above this threshold, and that there was a negligible number of prestellar sources. In our case, $\sim$ 20 per cent of protostellar sources and zero prestellar sources lie above this threshold.

\begin{figure}
    \centering
    \includegraphics[width=1\linewidth]{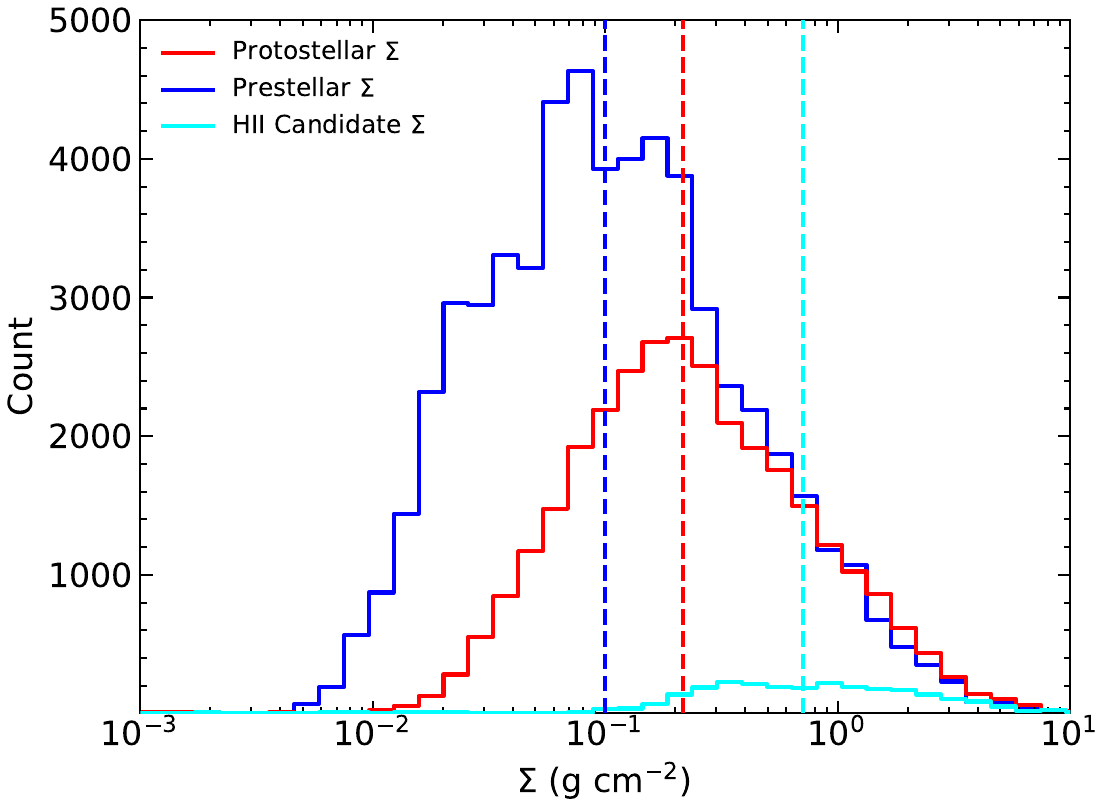}
    \caption{Same as in Fig.~\ref{fig:SigmaDist}, but for SCBD models.}
    \label{fig:SCBD_SigmaDist}
\end{figure}

Seen in Fig.~\ref{fig:SCBD_SigmaDist} are the corresponding surface density distributions for the SCBD models. While the median surface densities (0.10, 0.22, 0.71 g cm$^{-2}$ for prestellar, protostellar and \ion{H}{II} region candidates respectively) of the different evolutionary stages are larger when compared to the CBD distribution (see Fig.~\ref{fig:SigmaDist}), the general trend in increasing surface density remains unchanged. In this case, there are significantly more prestellar sources present. While the median surface density for prestellar and protostellar sources are similar to those from \citet{Elia2021}, the median \ion{H}{II} region surface density is significantly higher. This trend is not seen in the CSC-II catalogue, however a general trend of increasing surface density up to and including the UC\ion{H}{II} region stage is seen in high-mass BGPS clumps \citep{Svoboda2016}. 

It is important to note that the disagreement with \ion{H}{II} region surface densities may be due to how these samples are defined. \citet{Elia2021} used the threshold of $L/M=22.4$~L$_\odot$/M$_\odot$ to classify candidate \ion{H}{II} regions, finding that 60~per cent of their sample above this threshold were associated with radio emission from the WISE catalogue of \ion{H}{II} regions \citep{Anderson2014}. This shows that this conservative threshold likely overestimates the number of true \ion{H}{II} regions, allowing for instance externally heated, low-mass sources to fall into the same category, and  may explain the low median surface densities reported for their \ion{H}{II} region candidates. Regardless, while 14 per cent of all protostellar sources meet this criteria, 98 per cent of all SCBD sources with a high-mass ZAMS star lie above this $L/M$ threshold. 

\begin{figure}
    \centering
    \includegraphics[width=1\linewidth]{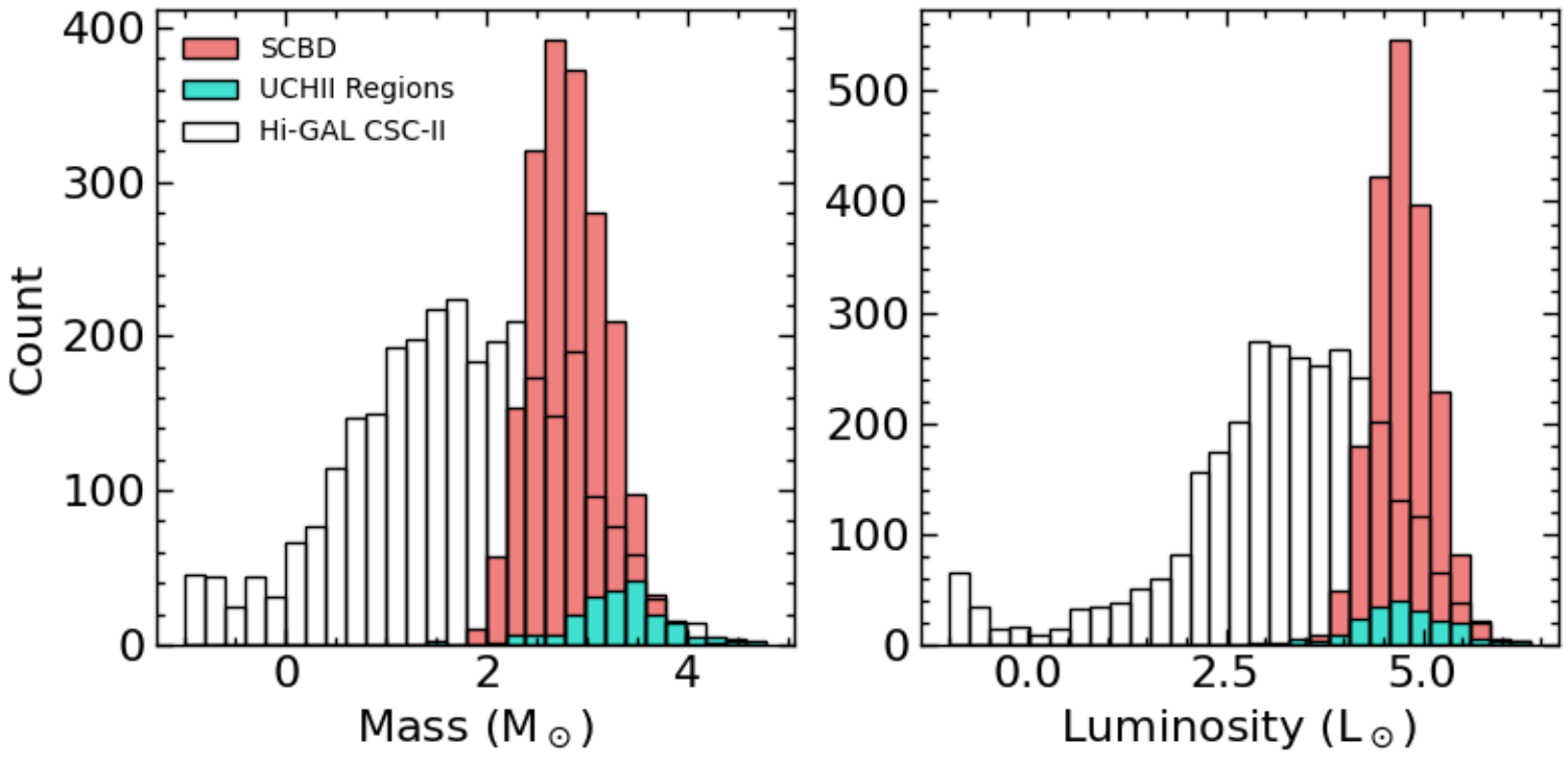}
    \caption{Distributions of mass and luminosity for \ion{H}{II} regions of varying definition. The UC\ion{H}{II} region sample is from \citet{Cesaroni2015}.}
    \label{fig:HII_Region_Def}
\end{figure}

On the other hand, the \citet{Cesaroni2015} sample of Hi-GAL clumps associated with CORNISH UC\ion{H}{II} regions span a much more narrow range of mass and luminosity. As seen in Fig.~\ref{fig:HII_Region_Def}, the sample of SCBD clumps classified as \ion{H}{II} region candidates agrees well with the observational sample. The subset of high-mass BGPS clumps associated with UC\ion{H}{II} regions from \citet{Svoboda2016} also show a trend of increased mass and surface densities relative to the less evolved stages, contrary to what is seen in the large sample of compact sources in \citet{Elia2017}. This demonstrates that the sample of SCBD clumps containing high-mass stars / \ion{H}{II} region candidates are more akin to observationally defined UC\ion{H}{II} regions than those met through $L/M$ thresholds. 

\section{Analysis}
\label{sec:Analysis}

\subsection{High-Mass Star Formation Surface Density Thresholds}

\subsubsection{Identification of Model Clumps Associated with High-Mass Star Formation}
As part of our analysis of the SCBD model outputs we wanted to identify SCBD clumps that have observable high-mass star formation tracers such as methanol masers or radio emission from \ion{H}{II} regions. The SCBD model does not do this directly, but it does provide useful information about the clump properties and stellar content that can be utilized. 

Average properties of observed regions exhibiting tracers of early high-mass star formation can be used to define these samples. For instance, high-mass ZAMS stars are classified in this paper as \ion{H}{II} region candidates. This is supported by the fact that nearly all of the SCBD clumps which contain high-mass ZAMS stars have $L/M > 22.4$ L$_\odot$/M$_\odot$, and have mass and luminosity distributions similar to true UC\ion{H}{II} regions (see Section~\ref{sec:SCBD_Compared_to_obs}). 

Given the large timesteps used in these models, when ZAMS stars appear ($10^6$-$10^7$ yr) it is safe to assume that one or more observable high-mass star formation tracers would also appear. For example, \citet{Haemmerl2016} found that the delayed onset of \ion{H}{II} region expansion is only on the order of $\sim 10^4$ yr. Additionally, masers have been found to trace the early evolutionary stages of star formation \citep{Breen2010} and can be found in dense ATLASGAL clumps, although the exact evolutionary sequence for maser emission is uncertain \citep{Billington2020}. 

For earlier stages we used results from \citet{Urquhart2013}, where they found Methanol Multibeam Survey associated clumps with luminosities $>10^3$ L$_\odot$ were associated with high-mass star formation. Those with lower luminosities were likely associated with intermediate-mass star formation due to their lower mass, and would require large star formation efficiencies to form high-mass stars. Given these observational results, SCBD models which contain high-mass ZAMS stars, or those containing high-mass YSOs ($>8$ M$_\odot$) with total luminosities in excess of $10^3$ L$_\odot$ are classified as candidate high-mass star-forming clumps with detectable tracers. 

We also developed a similar classification scheme for lower-mass model clumps. Following the results from \citet{Urquhart2013}, if a clump housing a maximum mass YSO ranging from 3 $\leq$ M$_\odot$ $<$ 8 has a luminosity $\leq$ $10^3$ L$_\odot$, it was classified as a candidate intermediate-mass star-forming clump with detectable tracers. Models which met this criterion but later went on to produce high-mass stars were excluded to limit the sample to true intermediate-mass star-forming regions. 

\subsubsection{Comparison with Commonly Used Thresholds}
\label{sec:SCBD_comp_with_thresh}
\begin{figure*}
    \centering
    \includegraphics[width=0.48\textwidth]{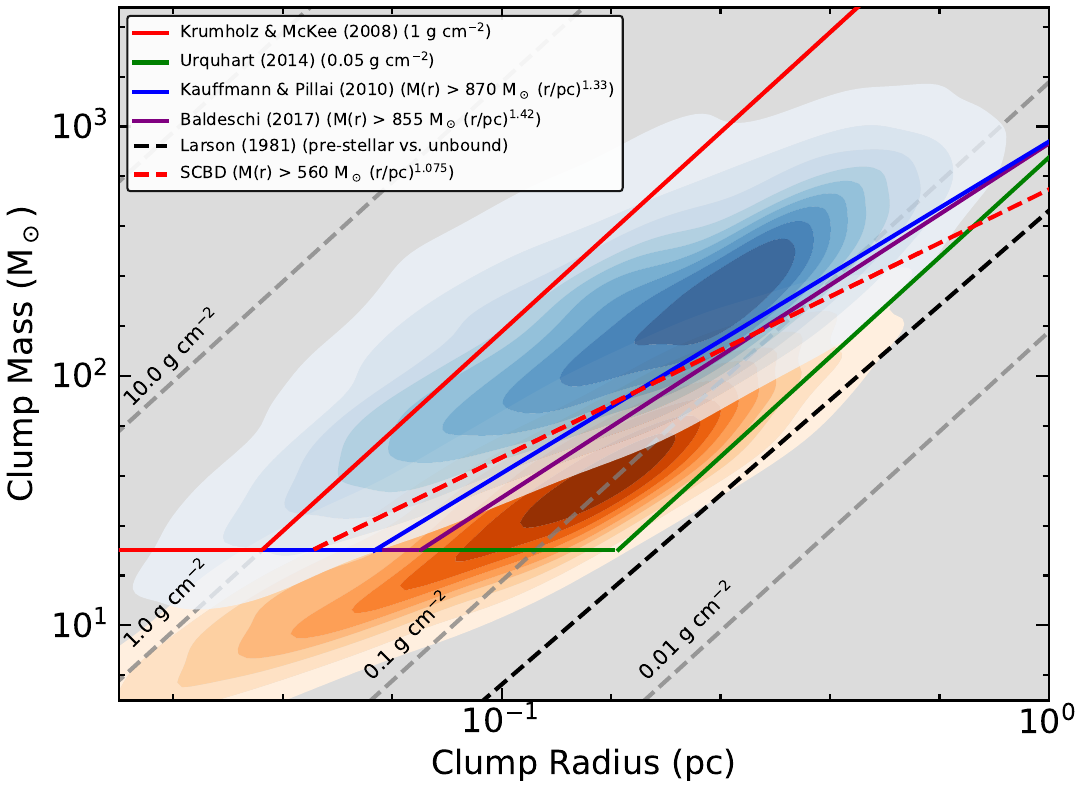}
    \hfill
    \includegraphics[width=0.48\textwidth]{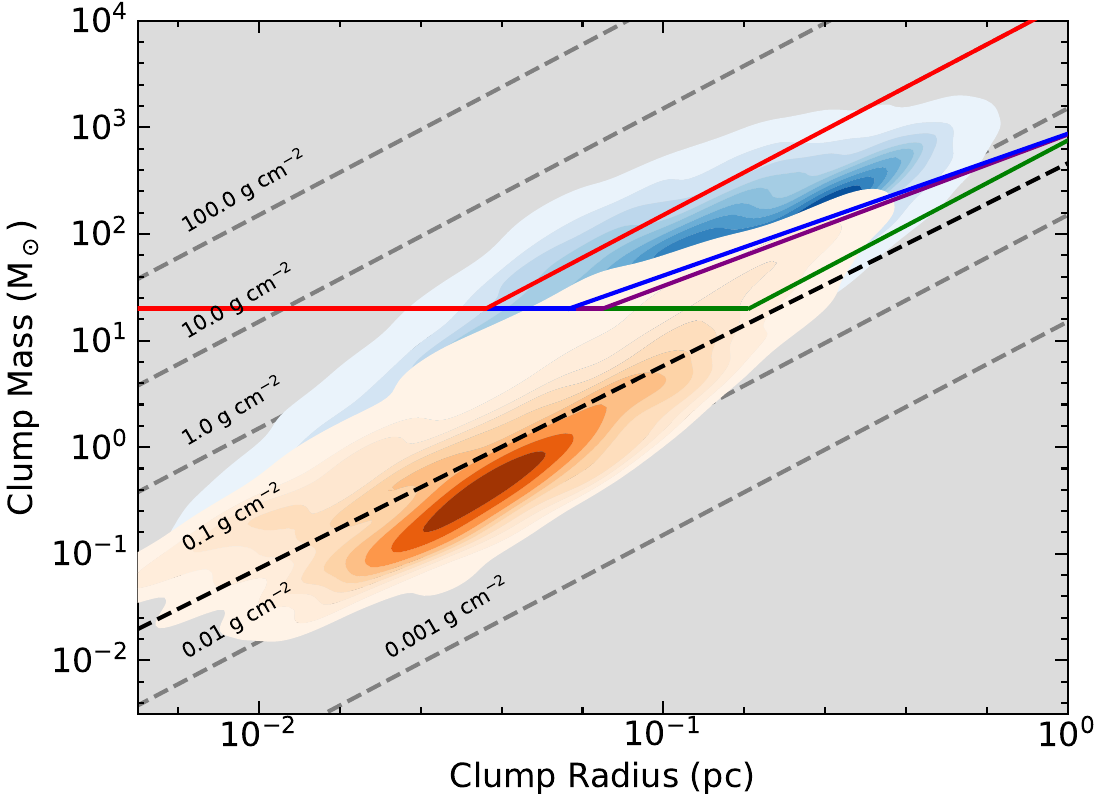}
    \caption{Left: Mass-Radius kernel density estimate (KDE, a representation of the data using a (2D) continuous probability density curve) plot from the SCBD model of candidate intermediate- (orange) and high-mass (blue) star-forming clumps with detectable tracers. High-mass clumps are defined as either those with a high-mass ZAMS stars, or containing a YSO mass $>$ 8 M$_\odot$ with $L>1000$ L$_\odot$. Intermediate-mass clumps are those that are either only ZAMS intermediate-mass stars or containing a YSO mass between 3 and 8 M$_\odot$, with $L<1000$ L$_\odot$. Commonly used thresholds associated with star formation are indicated in the legend (see text for details). The red dashed line is the fit to the SCBD data where only $\sim$10~per cent of high-mass clumps lie beneath the line, while simultaneously $\sim$10~per cent of intermediate-mass clumps lie above the line, which represents a threshold separating a majority of true intermediate versus high-mass star-forming clumps. This fit has the form $M(r)>558.5 (r/[pc]) ^{1.085}$. The horizontal line at $20$ M$_\odot$ is a lower limit for massive star formation assuming core-to-star conversion efficiency of 0.5-0.33 \citep{Baldeschi2017}. Right: The same plot all points defined by the final state of the clumps evolution. Blue points are those which contain a ZAMS, high-mass star at the end of the model evolution, while orange are those that contain a ZAMS, intermediate-mass star and no high-mass star. This panel covers a larger range than the left panel as it includes the future evolution of earlier evolutionary stages. Note the range of the axes in this panel is extended relative to the left panel, as the left panel is intended for comparison between commonly used thresholds.}
    \label{fig:MR_HM_IM}
\end{figure*}

Shown in the left-hand panel of Fig.~\ref{fig:MR_HM_IM} are the locations of the models classified as intermediate (orange) and high-mass (blue) star-forming regions at varying stages of evolution along with common surface density thresholds ($\Sigma_\text{th}$) values found in the literature. 

Quantile regression fits of the form $M(r) \sim a(r/[pc])^{b}$ were applied to the data such that 90~per cent of the high-mass points lay above the fit and separately 90~per cent of the intermediate-mass points lay beneath the fit. These fits were averaged together to form the red dashed line, $M(r) = 558.5(r/[pc])^{1.085}$ seen in the left-hand panel of Fig.~\ref{fig:MR_HM_IM}, representing a rough delineation between a majority of both the intermediate and high-mass model data points. The functional forms of the individual fits are $M(r) = 557(r/[pc])^{1.08}$ and $M(r) = 560(r/[pc])^{1.09}$ for the high-mass and intermediate-mass samples respectively.

For comparison, the observationally defined high-mass thresholds of 1.0 g cm$^{-2}$ \citep{Krumholz2008}, $M(r) = 870(r/[pc])^{1.33}$ \citep{KauffmannPillai2010ApJ} and $M(r) = 1282(r/[pc])^{1.42}$ \citep{Baldeschi2017} are shown in Fig.~\ref{fig:MR_HM_IM} as the red, blue and purple lines respectively. Additionally, the threshold for efficient star-formation \citep{Urquhart2013} and the threshold delineating prestellar versus unbound clouds \citep{Larson1981} are shown as the solid green and black-dashed lines. We see the model-derived threshold has a shallower slope than the \citet{KauffmannPillai2010ApJ} and \citet{Baldeschi2017} thresholds, allowing for both \textit{lower} surface densities at higher mass and \textit{higher} surface densities at lower mass to create high-mass stars. All of the high-mass points lie above the \citet{Urquhart2013} threshold for efficient high-mass star formation at 0.05 g cm$^{-2}$. Nearly 95~per cent of all points that pass above the 1.0 g cm$^{-2}$ threshold imposed by \citet{Krumholz2008} belong to the high-mass classification, with only 4.7~per cent being intermediate-mass. 

Our model-derived threshold, which is similar to the \citet{KauffmannPillai2010ApJ} and \citet{Baldeschi2017} observationally derived thresholds, does roughly separate the two regions containing the bulk of the intermediate versus high-mass star-forming clumps, however there is significant overlap. Our results underscore the fact that surface density thresholds should not be thought of in terms of a lower-limit for high-mass star formation but rather a dividing line where above it, a majority of clumps are actively forming high-mass stars. To further illustrate this, the right-hand panel of Fig.~\ref{fig:MR_HM_IM} shows all points for $\tau\leq\tau_\text{acc}$ colour coded by the final stage of their model evolution. The blue points represent all clumps where at the final timestep of their model there is at least one ZAMS high-mass star, and the orange points indicate those which eventually go on to form at least one ZAMS intermediate mass star without any high-mass stars. 

This highlights the issue addressed in this paper; namely that high-mass star formation thresholds are based on the current evolutionary stage of the clump and not on its final state. Without tracers of high-mass star formation, it becomes exceedingly difficult to predict the final cluster content. For example, consider a clump observed with 10~M$_\odot$ with a radius of 0.1~pc. The SCBD model predicts it \textit{could} evolve into either a low-, intermediate- or high-mass star-forming region. In the next section the SCBD models (classified as in the right hand panel of Fig.~\ref{fig:MR_HM_IM}) will be explored to see if any input or output parameters can be used to distinguish between them. 

\subsection{Applications}
As discussed in Section~\ref{sec:SCBD_comp_with_thresh}, surface density thresholds seen in the literature are able to distinguish between relatively evolved clumps containing high-mass stars and those only containing lower-mass stars. Without additional information, star-forming clumps beneath these thresholds could either be evolved intermediate- or low-mass star-forming regions, or younger high-mass star-forming regions still accumulating material. Therefore if $\Sigma_\text{th}$ is to be thought of as a \textit{minimum} surface density for clumps to eventually produce high-mass stars, it is likely to be lower than values reported in the literature. 

In a conveyor-belt style model, any star-forming region examined far enough back in its evolution will be essentially indistinguishable from any other. It would be useful to be able to differentiate between high-mass and not high-mass star-forming clumps at any point in their evolution, and not just when they are evolved enough to identify via tracers. As mentioned previously, the models we present do not include detailed radiative transfer calculations and important information such as maser lines or PAH emission. This information could prove beneficial in classifying relative evolutionary stages and providing more realistic SEDs for analysis, but is outside the scope of this paper.

\subsubsection{Differentiating Intermediate and High-Mass Star-Forming Regions}
The SCBD models have numerous properties that can be obtained from either the input parameters (such as $t_\text{sf}$, $M_\text{g,0}$), direct outputs from equation~(\ref{eq:Mgtinittau0}) and equation~(\ref{eq:Msinittau0}) ($M_g$ and $M_*$), or derived from the model outputs (e.g. luminosity, colours). We employ logistic regression (LR) as a statistical test in an attempt to differentiate models that go on to produce high-mass stars and those that do not. LR is a machine learning algorithm commonly used for binary classification, which predicts the probability of an outcome based on input variables. In this case, it is being used to predict whether a model goes on to form a high-mass star or not based on the input parameters discussed below. The parameters used in the context of LR are considered `features', and certain features are in principle directly observable or can be derived from observations. This includes mass, accretion rate, size, luminosity, and $\eta$, as well as age indicators such as $L_\text{bol}/M$, $L_\text{bol}/L_\text{submm}$ and various colours. Although some may be more difficult than others to accurately measure in real star-forming regions (such as $\eta$, see \citealt{Motte2018}), they are nonetheless considered here to determine the feasibility of classifying the SCBD models. 

The default {\sc sklearn} {\sc LogisticRegression} package \citep{Pedregosa2011} using the `lbfgs' (limited-memory Broyden-Fletcher-Goldfarb-Shanno) solver was used, where $\sim$2000 iterations yielded convergence. The class weighting was set to balanced, which applies more weight to less populated datasets, since the majority of the SCBD models were classified as eventually going on to form high-mass stars, as discussed in Section~\ref{sec:HMPreSClumps}. Given the large number of features that can be compared for both classifications, a recursive feature elimination (RFE) routine is implemented, which is used to identify the most important features in differentiating the two classes. The code returns a relative measure of the importance of each feature, quantified by a coefficient $\beta$ (note this is a different $\beta$ from what was used in Section~\ref{sec:ModifiedModels}). The term $e^\beta$ is the odds ratio representing the factor by which the odds of the outcome change for each one-unit increase in the given feature while holding other features constant.

The number of returned features was altered until the Receiver Operating Characteristic Area Under the Curve (ROC AUC) was maximised. This value ranges from 0 to 1 and quantifies how well the model can differentiate between the two classes. This resulted in five being the optimal number of features: gas mass, $[160/250]$ colour, surface density, $L_\text{bol}/M$ ratio, and total luminosity. These were then split to create a training set from 30 per cent of the data, and a testing set comprised of the remaining 70 per cent. This resulted in a mean ROC AUC of 0.868, with the general results presented in the first two lines of \autoref{tab:RFE_output}.

\begin{table}
    \centering
    \caption{Logistic regression outputs. The first two sections are for the entire SCBD dataset, where `observable' indicates that only features that are in principle observable or can be derived from observations are used, while the `all' category includes non-observable features. The bottom two sections are for a reduced dataset using only points beneath the \citet{Baldeschi2017} $\Sigma_\text{th}$.}
    \label{tab:RFE_output}
    \begin{tabular}{ccccc}
        \hline
        Class & Precision & Recall & F1-score & Support \\
        \hline
        HM (observable, full) & 0.98 & 0.74 & 0.85 & 25014 \\
        IM (observable, full) & 0.27 & 0.87  & 0.42 & 2748\\
        \hline
        HM (all, full) & 1.00 & 0.93 & 0.96 & 25014 \\
        IM (all, full) & 0.61 & 0.97  & 0.75 & 2748 \\
        \hline
        \hline
        HM (observable, $\Sigma_\text{th}$) & 0.95 & 0.74 & 0.83 & 15011 \\
        IM (observable, $\Sigma_\text{th}$) & 0.36 & 0.80  & 0.50 & 2719\\
        \hline
        HM (all, $\Sigma_\text{th}$) & 0.99 & 0.91 & 0.95 & 15011 \\
        IM (all, $\Sigma_\text{th}$) & 0.64 & 0.94  & 0.76 & 2719 \\
        \hline
    \end{tabular}
\end{table}

One way to set up the logistic regression model is to see if a given SCBD model will become a high-mass or \textit{not} high-mass star-forming region (i.e. the rows starting with HM). In the case of a high-mass SCBD model, if the logistic regression model predicted it as being not high-mass, that is flagged as a false negative (FN). If the SCBD model were not high-mass and the logistic regression model predicted it as being high-mass, that is flagged as a false positive (FP). It then follows that a not high-mass SCBD model correctly classified from the logistic regression as not high-mass would therefore be a true negative (TN). In this context, `precision' indicates how often the model is correct when predicting a SCBD model as being high-mass, and can be written as TP / (TP + FP). A high precision indicates few false positives, which means it infrequently misclassifies not high-mass as high-mass. `Recall' in this case measures how often SCBD high-mass models are accurately predicted from all true instances in the model. This can be written as TP / (TP + FN). A high recall indicates few false negatives, which means it infrequently misclassifies high-mass as not high mass. The above argument can also be written in terms of the intermediate-mass models by simply writing the statements in terms of intermediate- and not intermediate mass star-forming regions (i.e. the rows starting with IM). Finally the F1-score is the harmonic mean of precision and recall, which can be written as $2$(precision $\times$ recall)/(precision + recall), and support is simply the number of true intermediate- and high-mass samples. 

The first two sections of \autoref{tab:RFE_output} then indicate that although the LR model appears to do a adequate job at correctly identifying high-mass star-forming regions in the SCBD model 
(missing $\sim$26~per cent of true high-mass models), the trained regression model has many FPs when attempting to classify intermediate-mass SCBD models, which means it regularly misclassifies high-mass models as being intermediate-mass. The relatively high recall for the intermediate-mass sample indicates few FNs, which means the model rarely classified true intermediate-mass SCBD models as being high-mass. In general, if the model finds an intermediate-mass SCBD model, it does a good job at classifying it as intermediate-mass, but regularly misclassifies high-mass SCBD models as being intermediate-mass. Although this could be due to the overwhelmingly larger sample of high-mass clumps, the balanced class weighting applied greater weight to the intermediate-mass sample to offset this bias. 

Subsection~\ref{sec:SCBD_comp_with_thresh} showed that sufficiently evolved clumps capable of being observationally classified as high-mass were generally found above commonly used $\Sigma_\text{th}$. However, one of our goals was to determine if less evolved high-mass star-forming regions beneath these thresholds could be distinguished from those going on to form intermediate-mass star-forming regions. For instance, when considering SCBD clumps beneath the \citet{Baldeschi2017} $\Sigma_\text{th}$, the regression results are similar to those using the full sample if not slightly improved, which is seen in the bottom two sections of \autoref{tab:RFE_output}, labelled as $\Sigma_\text{th}$.

If all features are included regardless of observability (including, for instance, the effective $M_\text{g,0}$ and $f_M$), the returned features become M$_{{g,0}_\text{eff}}$, $\eta$, L$_\text{bol}$, t$_\text{sf}$ and $f_M$. This results in an ROC AUC of 0.986, with general results presented in two lines beneath those using the observed dataset of \autoref{tab:RFE_output}. Even though the model continues to have difficulty in correctly identifying intermediate-mass star-forming regions, it performs significantly better than with the prior set of features. 

The effective $M_\text{g,0}$ was found to be the most important feature by a large margin. Similarly, \citet{Semadini2018} concluded from their models that the maximum mass ever attained by a cloud, equivalent to the total mass involved in the history of the star-forming region, is the main parameter defining a given model's evolution. In this case, the $\beta$ coefficient of $M_\text{g,0}$ is $\sim$3.5 times larger than the $\beta$ coefficient of the next most important feature, $\eta$. That $\eta$ is the second most important feature is unsurprising, as research has shown that the balance between the gravitational strength and stellar feedback is a key parameter in setting cluster masses (for example, see \citealt{Chen2021, Grudic2021, Semadeni2024_delay}). It is important to note that the features were standardised for this analysis, meaning the relatively large scale of $M_\text{g,0}$ values when compared to $\eta$ does not effect the resulting $\beta$ coefficients.

In order to apply this to real observations, knowledge of the total amount of material to ever enter the star-forming region (between $\tau_0$ and $\tau_\text{acc}$), the initial gas mass prior to the onset of star formation (or the maximum prestellar gas mass), total star-formation timescale and/or $\eta$ must also be known. This is not surprising, as this tells us that if more information about the clump related to its entire formation history is available, differentiating its end result becomes a much more simple task. 

Every model, regardless of the initial $f_M$, forms a high-mass star above an effective $M_\text{g,0}$ of $\sim10^{3.5}$ M$_\odot$, while every model beneath an effective $M_\text{g,0}$ of $\sim10^{2.3}$ M$_\odot$ produces intermediate-mass star-forming regions or lower. To test if this is unique to the SCBD model, the same procedure was performed on the original CBD model as well. The CBD model similarly produced high-mass stars for all models with $M_\text{g,0}$ $\gtrsim 10^{3.2}$~M$_\odot$, while those beneath $\sim10^{2.2}$~M$_\odot$ produced intermediate-mass (or lower) star-forming clumps. This implies a global cutoff for the conveyor-belt mechanism, where given sufficient amounts of material any model will eventually go on to produce a high-mass star-forming region, regardless of other features. 

The use of LR analysis in this work has shown to be able to discriminate between intermediate- and high-mass star forming regions outside of certain parameter ranges. Extending the current analysis to explicitly evaluate pairwise combinations of mass and other evolutionary indicators, similar to the approach taken by \citet{Semadini2018}, represents an interesting area of future work.

\subsubsection{Strategies for Predicting the End State of Cluster Formation}
Even though $M_\text{g,0}$ has been shown to be useful in distinguishing between the models classified by their eventual stellar content, this parameter is not directly observable. However, current models of star formation suggest potential ways forward.

For instance, the `filaments to clusters' paradigm proposed by \citet{Kumar2020} outlines a scenario where flow-driven filaments overlap to form a junction called a hub. While low-mass star formation can occur within the filaments independently, the enhanced gravitational potential from the hub triggers longitudinal flows, bringing additional matter towards the hub. The hub can then go on to fragment and produce a star cluster(s), with radiative pressure and ionisation feedback escaping through the inter-filamentary cavities. These HFSs are observed to exist as multi-scale, hierarchical structures that range in evolutionary stage from quiescent to infrared bright hosting UC\ion{H}{II} regions (e.g. \citealt{Zhou2023,Liu2023,Yang2023,Bhadari2025}). 

\citet{Kumar2020} used SPIRE 250 $\mu$m images to identify filaments, where the filament skeletons could be used as masks to obtain column-density maps. Applying this procedure to the Hi-GAL CSC-I \citep{Elia2017} clumps, they found large samples of candidate prestellar and protostellar HFSs. They also found a large fraction of their models were classified as non-hubs, where $<3$ interconnecting filaments were found adjoined to the central clump. 

If one were to consider a low-mass prestellar clump classified as a HFS candidate adjoined to multiple filaments, the primary factor determining the future evolution of the clump would be how much material the clump continued to accumulate through the remainder of its life ($M_\text{g,0}-M_\text{init}$). Assuming the interconnected filaments acted as reservoirs to the central hub, the accretion rate and total reservoir mass may be used to estimate an upper limit to how much material could be accumulated. An isolated, prestellar $\sim$20-40 M$_\odot$ clump likely would not accumulate enough material to produce high-mass stars, whereas one connected to accreting filaments would be much more likely. This type of analysis would require high-resolution observations of non-hubs and HFS candidates over a wide range of masses to create a statistically significant sample. As a starting point, confirmed HFSs found in the literature combined with $>3\times10^3$ candidates from \citet{Kumar2020} could be used.

Another related avenue forward involves looking at the large scale structure associated with star-forming regions. For example, \citet{Lada2010} determined that the number of YSOs in a given molecular cloud was related to the cloud mass above $A_{K}\sim0.8$ mag, and the SFR was related to the mass surface density above $116$ M$_\odot$ pc$^{-2}$. Similarly \citet{Jiao2025} found that the dense gas mass above $A_V>8$ or the mass derived from column densities above a threshold density set by the turnover point in column density probability distribution functions (N-PDFs) correlate well with the SFR. The dense gas mass has also been used in studies such as \citet{Semadini2018} to determine cloud ages through interpolation of a grid of cloud models, allowing them to pinpoint a cloud along its evolutionary path. \citet{Howard2018} stated that in the case of globular clusters, the final mass of the largest young massive cluster (formed through accretion and cluster mergers) corresponds to the mass of the parent giant molecular cloud. 

These trends between large scale properties of molecular clouds (total cloud mass, dense gas mass) and stellar content/formation rate motivate studies looking for differences between low- and high-mass star-forming regions. \citet{Jiao2025_2} found trends between the most massive core mass and dense gas mass (mass above N-PDF log-normal and power-law turnover) for samples of low- and high-mass star-forming regions, suggesting a potential correlation between bound gas mass, maximum core mass, maximum stellar mass, the overall cluster mass, and the SFR in molecular clouds. Analyses of N-PDFs for a sample of low- and high-mass star-forming regions in \citet{Schneider2022} indicate the peak $A_V$ and deviation points in $A_V$ from the low density, log-normal portion and the dense gas power-law portion of the N-PDFs show an increasing trend with cloud type. These differences imply a potential transition point where intermediate-mass star-forming regions could be found, which may be identified through analysis of the large scale cloud or reservoir. 

Recently, large samples of candidate intermediate-mass star-forming regions have become available in the literature through colour analysis and catalogue cross-matching (e.g. \citealt{Lundquist2014, Wolf-Chase2021ApJ, Devine2026}). Although many of these sources may be true intermediate-mass star-forming regions, additional confirmation is necessary to rule out future evolution into UC\ion{H}{II} regions. Given that low- and high-mass star-forming regions appear to form in differing environments as seen from their column density distributions and SFRs, these metrics may be used to identify regions likely to go on to form high-mass stars versus those that will not.

\section{Summary and Conclusion}
\label{sec:Conclusions}
In this paper, synthetic clumps and their resulting young star clusters were created using the conveyor belt model with dispersal post-accretion (CBD) \citep{KrumholzMcKee2020} to investigate the evolution of their surface densities, and compare them to commonly used surface-density thresholds. These clumps were created following the procedure outlined by \citet{Molinari2019}, where gaseous clumps were populated with YSOs taken from \citet{Robitaille2006} according to the total stellar mass at each time step using a Kroupa IMF \citep{Kroupa2001}. The SED of the clump was constructed by summing the dust emission from the gaseous clump and the dust-extincted YSOs. 

These models resulted in evolutionary tracks ($L$ vs. $M$) differing from those found in the literature \citep{Molinari2019}. This difference is due to the accelerating accretion rate yielding an increase in clump mass with time, as opposed to other models where pre-assembled clumps lose mass to star production. Although the general locations of protostellar and ZAMS-populated high-mass clumps agreed with observations (e.g. mass, luminosity, radius, surface density), the sample of prestellar clumps disagreed with observations.

This prompted the creation of the seeded CBD, or SCBD model, in which clumps have some initial non-zero gas mass prior to the onset of star formation. The increased accretion rate relative to the non-seeded model resulted in the total amount of material to enter the star-forming regions over their lifetime to be scaled by a factor of (1+$\beta$)$^{(p+1)}$. This model produced results which not only agreed with ATLASGAL $\epsilon_\text{ff}$ distributions, age distributions from young clusters and total predicted SFR, but was also able to produce distributions of high-mass, prestellar clumps similar to those seen in large clump catalogues.

A subset of models was selected assuming the clumps could be observationally classified as candidate intermediate- and high-mass star-forming clumps. This sample, when plotted on a mass-radius diagram, showed a clear distinction between a majority of those classified as high- versus intermediate-mass. The dividing line between the two groups found through quantile regression resulted in a $\Sigma_\text{th}$ of the form $M(r) = 558.5(r/[pc])^{1.085}$. This model-derived threshold is in good agreement with observationally-derived $\Sigma_\text{th}$ from \citet{KauffmannPillai2010ApJ} and \citet{Baldeschi2017}.

The evolutionary tracks presented in this paper suggest that many star-forming regions that go on to produce high-mass stars at some point fall beneath commonly used thresholds for high-mass star formation. In an attempt to discern between young clumps lacking signs of star formation which go on to form high-mass stars or not, logistic regression was implemented. When observable parameters such as mass, radius, luminosity, and colour were included as features, the model struggled to correctly identify intermediate-mass star-forming regions. However, when including additional features such as $M_\text{g,0}$ and $f_M$, the model performed significantly better, with the former being the most important feature in distinguishing between the two. Although information regarding the star formation history can be difficult to obtain, these results suggest that regions that reside in or near large reservoirs (such as HFSs) are more likely to form high-mass stars. Additionally, environmental analysis of regions forming low-, intermediate- or high-mass stars may be pertinent in determining the future evolution of young star-forming regions, as well as locating regions housing the precursors to high-mass stars. 

\section{Acknowledgements}
The authors would like to thank Enrique Vázquez-Semadeni for providing very useful and constructive feedback which helped improve the quality of this manuscript. The authors acknowledge partial support for this research from NSF grant No. 2307806.

\section*{Data Availability}
The derived data generated in this research will be shared on reasonable request to the corresponding author.

\bibliographystyle{mnras}
\bibliography{bib}

\appendix

\section{Alternate Evolutionary Classification}
\label{appendix:alt_evol_class}
In Section~\ref{sec:init_results} and throughout the remainder of the paper, the model definitions of evolutionary classes were used, which assumed that prestellar clumps were those with total stellar masses $<$0.1 M$_\odot$, or equivalently those without any populated YSOs. Protostellar sources were those with stellar masses $\geq$0.1 M$_\odot$, and \ion{H}{II} region candidates were those housing high-mass ZAMS stars (those with spectral type earlier than B3).

While these definitions applied to the SCBD model resulted in distributions of properties (luminosity, mass, surface density) that agreed well with observations, there exist alternate ways to define these datasets. For instance, observations tend to differentiate prestellar and protostellar sources by a detection (or lack thereof) at 70~$\mu$m, an indicator of warm dust surrounding YSOs (\citealt{Svoboda2016, Elia2017}). Additionally, \citet{Elia2017,Elia2021} used $L_\text{bol}$/$M>$ 22.4 L$_\odot$/M$_\odot$ as a conservative threshold for protostellar sources without radio emission that could be \ion{H}{II} region candidates. 

When using the model classifications, the classification scheme was independent of distance. When using the observational classifications, increasing the distance modifies the SEDs directly, namely through interstellar extinction and the reduction of the flux density through the inverse square law. Therefore by moving the source sufficiently far away, the flux density at 70~$\mu$m could fall beneath sensitivity limits resulting in a misclassification. To account for this, the models (which are all originally `observed' at 1.0 kpc) were randomly moved to distances between 1.0 and 10.0 kpc. Depending on the chosen distance, the inverse square law along with an $A_V$ of 1.0 mag kpc$^{-1}$ interstellar extinction \citep{Whittet2022} was applied. 

\begin{figure}
    \centering
    \includegraphics[width=1\linewidth]{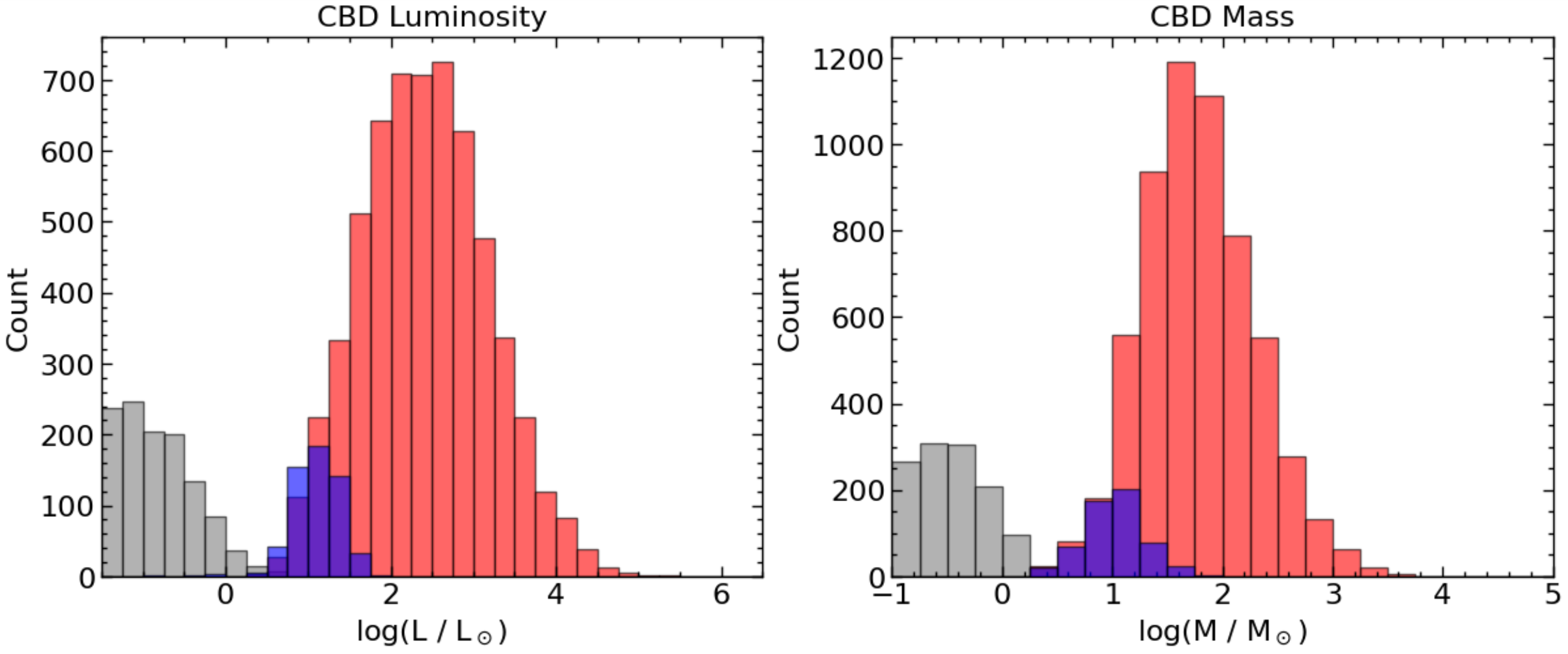}
    \caption{Luminosity and mass distributions of prestellar (blue), protostellar (red), and prestellar non-clumps (grey) using the  observational classification on the CBD model. }
    \label{fig:CBD_altdef_ML}
\end{figure}

\begin{figure}
    \centering
    \includegraphics[width=1\linewidth]{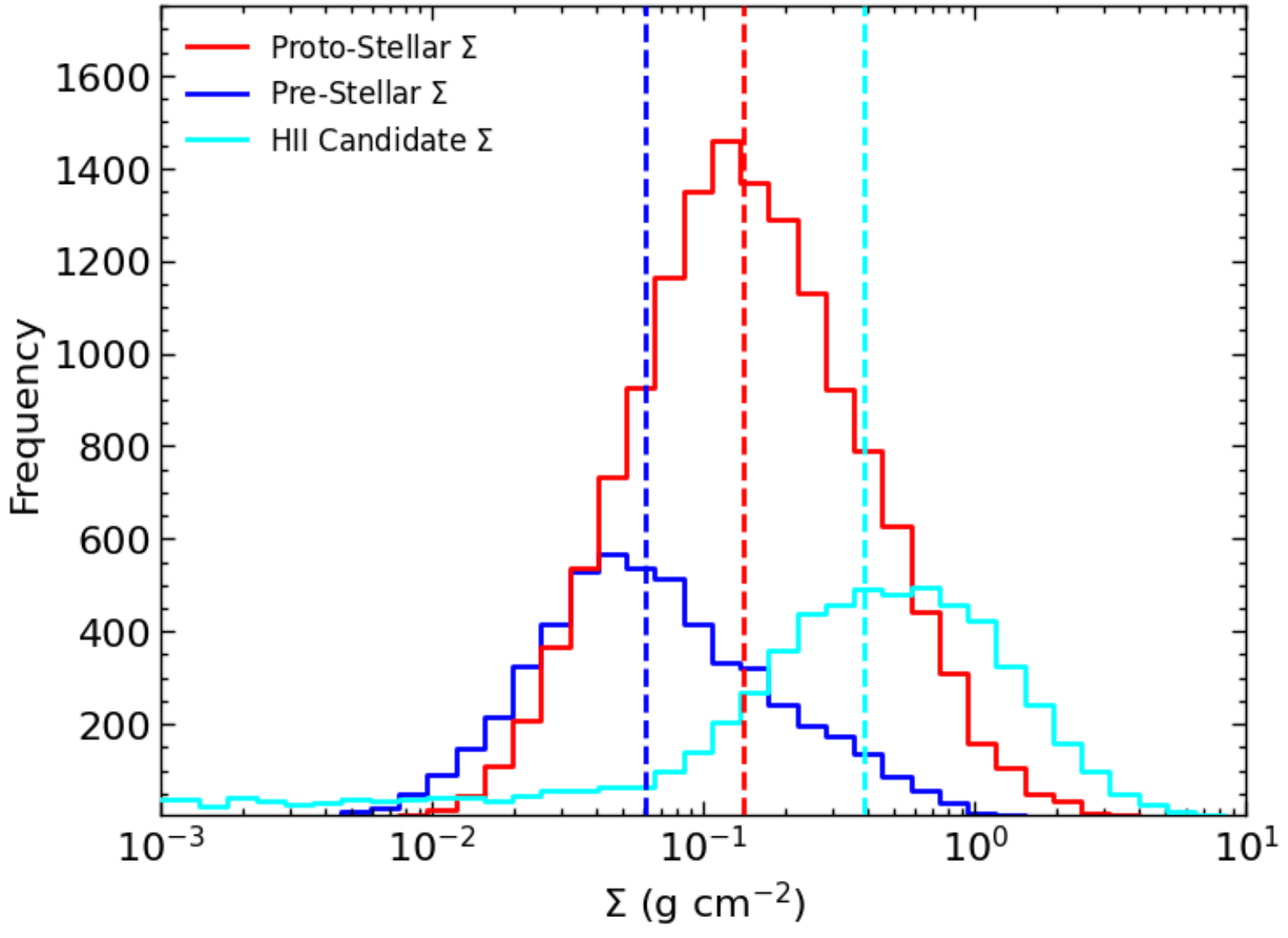}
    \caption{Surface density distribution of the CBD model using observational classifications.}
    \label{fig:CBD_Sigma_Alt}
\end{figure}

\begin{figure}
    \centering
    \includegraphics[width=1\linewidth]{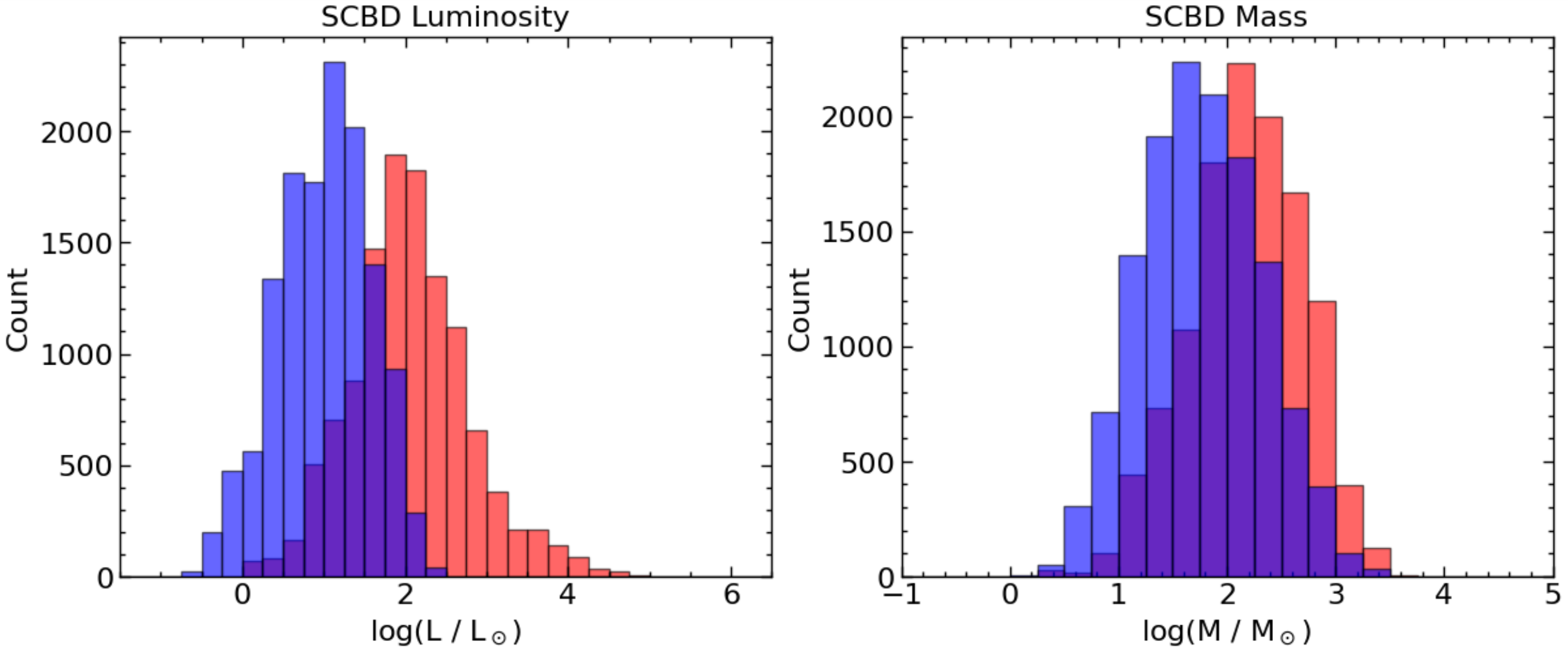}
    \caption{Same as in Figure B1, but for the observationally classified SCBD model.}
    \label{fig:SCBD_altdef_ML}
\end{figure}

\begin{figure}
    \centering
    \includegraphics[width=1\linewidth]{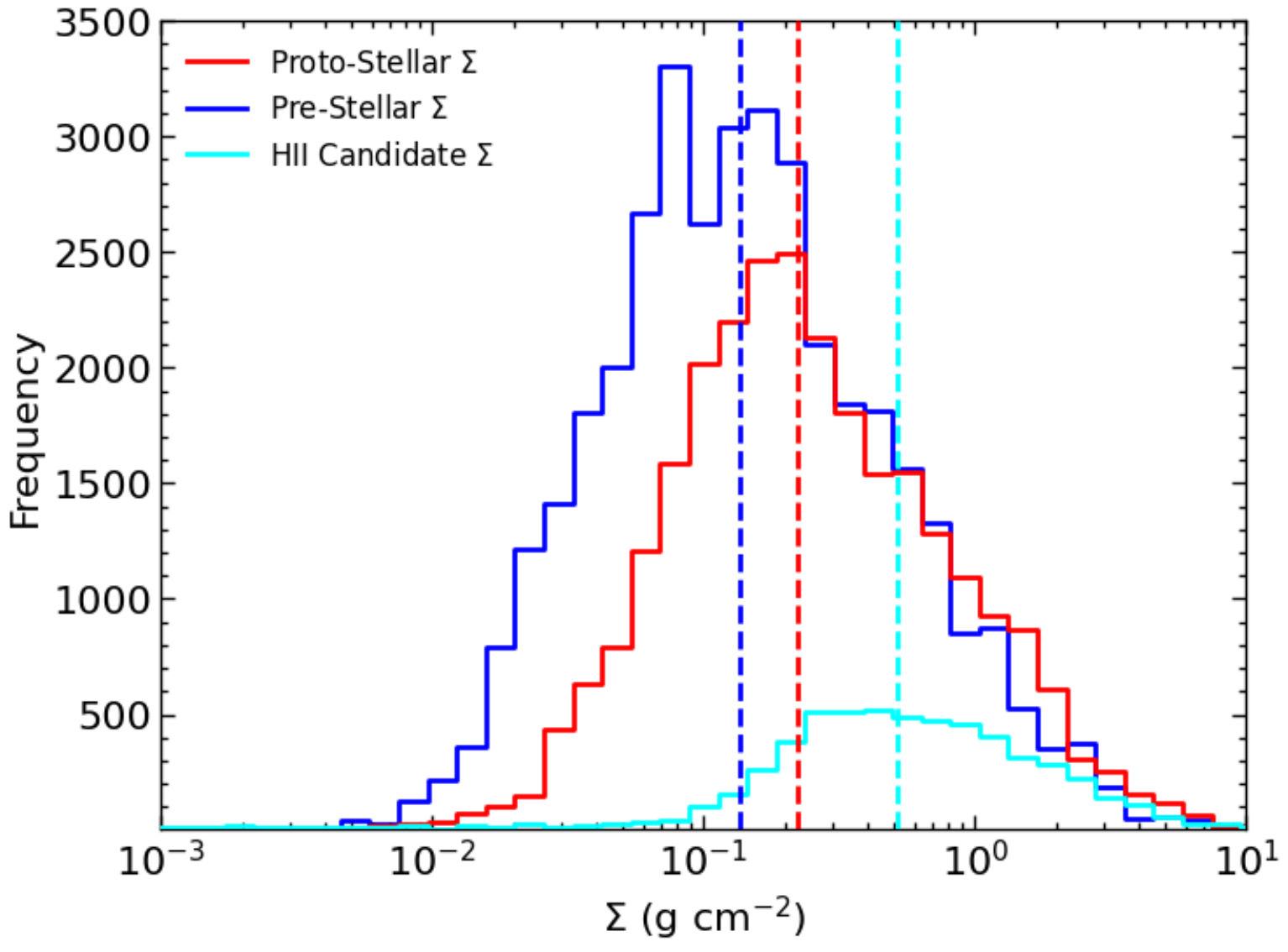}
    \caption{Surface density distribution of the SCBD model using observational classifications.}
    \label{fig:SCBD_Sigma_Alt}
\end{figure}

Shown in Fig.~\ref{fig:CBD_altdef_ML} and Fig.~\ref{fig:CBD_Sigma_Alt} are the resulting luminosity, mass and surface density distributions after using this alternate definition. As can be seen in Fig.~\ref{fig:CBD_altdef_ML}, defining the CBD model clumps this way allows the model to produce few prestellar clumps, however they are all low-mass ($<70 \;$M$_\odot$), and still fail to reproduce the results from observations seen in  Fig.~\ref{fig:age_fits}. Additionally, only $\sim$1-2~per cent of the sources on average contain $<$0.1 M$_\odot$ of stars, or are actually true prestellar clumps. This implies that either a large majority of observed clumps are false-positives (no detection at 70~$\mu$m is due to distance bias), or this model is unable to produce high-mass prestellar clumps regardless of the classification scheme employed. The results for the SCBD model seen in Fig.~\ref{fig:SCBD_altdef_ML} and Fig.~\ref{fig:SCBD_Sigma_Alt} show little difference in comparison with Fig.~\ref{fig:MinitLvsM}, with the only differences mainly being the relative number of sources classified as prestellar versus protostellar as discussed above. Finally, besides the relative fraction of sources in each classification, the surface density distributions for both models again remain unchanged.

\section{Derivation of the SCBD model}
\label{appendix:deriv_SCBD}
Presented here is the derivation for the functional form of the alternate model used in Section~\ref{sec:HMPreSClumps}. In this version, the formalism for the original conveyor belt model from \citet{KrumholzMcKee2020} is followed; however the star-forming clump is allowed to accrete material \textit{prior} to the onset of star formation, creating a reservoir that then goes on to form stars after `sufficient' material has been accreted. This amount is arbitrary, but allows for the formation of high-mass prestellar clumps prior to the onset of star formation. This is accomplished by shifting the accretion to start at an earlier time $\tau_0$. This modifies the accretion term seen in equation~(\ref{eq:AccRate}), in turn becoming
\begin{equation}
    \dot M_\text{acc} = H(\tau_\text{acc}-\tau)(p+1)\frac{M_\text{g,0}}{\tau_\text{acc}^{p+1}}\big(\tau + \tau_0\big)^p,
\end{equation}
such that at $\tau$ = 0, the accretion is non-zero and represents accretion due to the pre-assembled clump of mass $M_\text{init}$. Here $\dot M_\text{acc}$ is with respect to $\tau$. This is then used to solve the ODEs present in equation~(\ref{eq:GenFramework}). The differential form of the gas mass at $\tau < \tau_\text{acc}$ is now
\begin{align}
        \frac{dM_g}{d\tau} + M_g = (p+1)\frac{M_\text{g,0}}{\tau_\text{acc}^{p+1}}\big(\tau + \tau_0\big)^p,
\end{align}
which can be solved through the integrating factor technique, where I = $e^\tau$. Using the boundary conditions at $\tau'$=0 and $\tau'$=$\tau$, this leads to
\begin{align}
    M_g(\tau) - M_\text{init} = (p+1)\frac{M_\text{g,0}}{\tau_\text{acc}^{p+1}}e^{-\tau}\int_0^\tau\big(\tau' + \tau_0\big)^pe^{\tau} d\tau'.
\end{align}
The RHS can be simplified by performing a u-substitution, where u = $\tau' + \tau_0$, replacing it with
\begin{align}
    &= (p+1)\frac{M_\text{g,0}}{\tau_\text{acc}^{p+1}}e^{-\tau-\tau_0}\int_{\tau_0}^{\tau+\tau_0}u^pe^{u} du \\
    &= (p+1)\frac{M_\text{g,0}}{\tau_\text{acc}^{p+1}}e^{-\tau-\tau_0}\Big[\int_{0}^{\tau+\tau_0}u^pe^{u} du - \int_{0}^{\tau_0}u^pe^{u} du\Big].
\end{align}
From equation (9) in KM20 using their definition of \textit{g}, it can be seen that
\begin{align}
    \int_0^A x^pe^xdx = \frac{g(A,p+1)e^A}{(p+1)},
\end{align}
which allows the gas mass to be re-written as
\begin{align}
    M_g(\tau) = M_\text{init}e^{-\tau} + \frac{M_\text{g,0}}{\tau_\text{acc}^{p+1}}\Big[g(\tau+\tau_0, p+1)-g(\tau_0, p+1)e^{-\tau}\Big].
\end{align}
Using this and the second equation of equation~(\ref{eq:GenFramework}), the stellar mass can be derived from 
\begin{align}
    \frac{dM_*(\tau)}{d\tau} &= \frac{1}{(1+\eta)} M_g \\ 
    &= \frac{1}{(1+\eta)}\Big(M_\text{init}e^{-\tau} + \\ & \qquad\qquad \frac{M_\text{g,0}}{\tau_\text{acc}^{p+1}}\Big[g(\tau+\tau_0, p+1)-g(\tau_0, p+1)e^{-\tau}\Big]\Big).
\end{align}
Through direct integration of $\tau$, the stellar mass can be written as
\begin{align}
    M_*(\tau) = \frac{1}{1+\eta} \Bigg[&M_\text{init}(1 - e^{-\tau}) + \\
    & \frac{M_{g,0}}{\tau_\text{acc}^{p+1}} \Bigg(\frac{g(\tau+\tau_0, p+2)}{p+2}-\\
    & \qquad \frac{g(\tau_0, p+2)}{p+2} - g(\tau, p+1)(1 - e^{-\tau})\Bigg)\Bigg].
\end{align}
Similarly, post-$\tau_\text{acc}$ the gas mass can be found by setting $\dot{M_\text{acc}}$ to zero in the first equation of equation~(\ref{eq:GenFramework}), which leads to  
\begin{align}
    M_g(\tau) = M_g(\tau_\text{acc})e^{-\phi_d(\tau-\tau_\text{acc})}.
\end{align}
This form of the gas mass can now be substituted into the second equation of equation~(\ref{eq:GenFramework}), resulting in
\begin{align}
    &M_*(\tau_\text{acc}) + \frac{1}{1+\eta} [M_g(\tau_\text{acc})(1-e^{-\phi_d(\tau-\tau_\text{acc}))})].
\end{align}

\bsp	
\label{lastpage}
\end{document}